\documentclass[prd,preprint,showpacs,groupedaddress]{revtex4-1}
\usepackage{amsmath}
\usepackage{amsfonts}
\usepackage{amssymb}
\usepackage{geometry}
\usepackage{graphicx}
\usepackage{natbib}
\usepackage[english]{babel}
\usepackage{graphicx}
\usepackage{subfigure}
\usepackage{epstopdf}
\usepackage{caption}
\usepackage{multirow}
\usepackage{indentfirst}

\usepackage{diagbox}
\usepackage{mathrsfs}
\usepackage{booktabs}
\usepackage[colorlinks,linkcolor=blue,anchorcolor=,citecolor=blue]{hyperref}
\usepackage{amsthm}
\usepackage{float}
\numberwithin{equation}{section}
\numberwithin{figure}{section}
\numberwithin{table}{section}

\begin{document}
	\title{Optical properties of Euler-Heisenberg black hole in the Cold Dark Matter Halo}
	\author{Lei You}
	\affiliation{Lanzhou Center for Theoretical Physics, Key Laboratory of Theoretical Physics of Gansu Province, Lanzhou University, Lanzhou, Gansu 730000, China}
	
	\author{Rui-bo Wang}
	\affiliation{Lanzhou Center for Theoretical Physics, Key Laboratory of Theoretical Physics of Gansu Province, Lanzhou University, Lanzhou, Gansu 730000, China}
	
	\author{Shi-Jie Ma}
	\affiliation{Lanzhou Center for Theoretical Physics, Key Laboratory of Theoretical Physics of Gansu Province, Lanzhou University, Lanzhou, Gansu 730000, China}
	
	\author{Jian-Bo Deng}
	\affiliation{Lanzhou Center for Theoretical Physics, Key Laboratory of Theoretical Physics of Gansu Province, Lanzhou University, Lanzhou, Gansu 730000, China}
	
	\author{Xian-Ru Hu}
	\email[Email: ]{huxianru@lzu.edu.cn}
	\affiliation{Lanzhou Center for Theoretical Physics, Key Laboratory of Theoretical Physics of Gansu Province, Lanzhou University, Lanzhou, Gansu 730000, China}
	
	\date{\today}
	\begin{abstract}
		The optical properties of Euler-Heisenberg (EH) black hole (BH) surrounded by Cold Dark Matter (CDM) halo are investigated. By changing BH's parameters, we found that the radius of horizon $r_{h}$ and radius of photon sphere $r_{ph}$ will transparently increase as CDM halo parameters $R$ and $\rho$ increase. To show the influence of CDM halo on the BH's optical characteristics, we took two sets of $R$ and $\rho$ with prominent differences and plot the first four orders of images for thin accretion disk with different angle of inclination $\theta$ of observer. The images with light intensity distributions using Novikov-Thorne (N-T) model are also derived, as well as the effective potential, photon orbits. Especially, analysis of intersection behaviors between photon trajectories with different impact parameters and circular time-like orbits in accretion disk will help better understand the image of thin accretion disk. Our results showed that CDM halo will make BH become more larger and dimmer distinctly.
		
	\end{abstract}
	
	\maketitle
	\section{Introduction}
	    Since the inception of General Relativity (GR), theoretical studies concerning black holes have spanned nearly a century, yielding significant achievements. However, it wasn't until 2019 that the Event Horizon Telescope (EHT) captured the first-ever image of the supermassive BH at the center of the Messier 87*(M87*) galaxy \cite{M87}. Moreover, the EHT discovered that the magnetic field structure of the accretion disk is consistent with predictions from the General Relativity Magnetohydrodynamics (GRMHD) model \cite{M87CXJ1,M87CXJ2}. Not long after, the EHT obtained an image of the black hole in the Milky Way galaxy's Sagittarius A*(SgrA*), and it was found that the size of the ring around SgrA* is consistent with GR's prediction within $10\%$ \cite{SA1,SA2,SA3,SA4,SA5,SA6}.In this image, we can clearly observe a bright ring-like structure outside the shadow of the BH, which corresponds to the emitted light from the BH's accretion disk. Therefore, research on BH accretion disks will contribute to our understanding of the properties of the BH surrounded by them and validate existing theoretical models.
	    
	    The accretion disk, a thin, disk-shaped structure composed of high-energy matter, orbits around massive celestial bodies \cite{BHM}. BH, as supermassive celestial entities, exert immense gravitational forces that attract surrounding matter to form their accretion disks. Research on BH accretion disks dates back to the 1970s, with Shakura and Sunyaev proposing the Shakura-Sunyaev model, which conceptualizes the accretion disk as a disk composed of viscous gas rotating around a central BH or star \cite{BXJP}. Page analyzed the time-averaged structure of a disk of material accreting onto a BH \cite{tav}, while Luminet employed elliptic integral methods to study accretion disk imaging of Schwarzschild BH \cite{F1}. Cunningham investigated the impact of different viewing angles on accretion disk imaging of Kerr black holes \cite{KEXJP}. To date, research on BH accretion disks has made significant strides and has become quite mature in scientific inquiry.
	    
	    The coupling between nonlinear electrodynamics and gravity has been a subject of considerable interest. One of the most prominent examples is the gravitational Born-Infeld (BI) theory \cite{BI}. The discoveries in string theory and D-brane physics have led to the emergence of Abelian and non-Abelian class BI-like Lagrangians in the low-energy limit \cite{XL1,XL2,XL3}, reigniting interest in such nonlinear behaviors. The effective Lagrangian for nonlinear electromagnetic fields was proposed by Euler and Heisenberg based on Dirac's positron theory \cite{EH1936}. This approach predicted one-loop corrections in quantum electrodynamics (QED) and investigated vacuum polarization in QED. This novel method enabled nonlinear electrodynamics(NLED) models to explain the expansion of the universe at its inception \cite{US}. In \cite{EHJ}, the EH BH solution was derived by examining the one-loop Lagrangian density associated with the Einstein field equations.
	    
	    Meanwhile, dark matter remains a crucial subject in modern physics, as it has yet to be fully understood. Dark matter constitutes approximately 85$\%$ of the total matter in the universe \cite{CDM3,CDM2,CDM1}, yet it does not participate in electromagnetic interactions. Therefore, researchers can only study it through its gravitational effects. Numerous models of dark matter have been proposed, among which the CDM model, proposed by Peebles \cite{CDM4}, is one of them. 'Cold' refers to the fact that the particles of this type of dark matter move at velocities much lower than the speed of light. It is precisely because of this characteristic that it plays a crucial role in the large-scale evolution of the universe. Currently, observations of the large-scale structure of the universe are broadly consistent with predictions from the CDM model. Given this, studying the influence of CDM on BH will help us further understand the nature of dark matter \cite{CDMT}.
	    
	    Our work analyzed the imaging of the accretion disk surrounding an EH BH enveloped by CDM halos. Our paper is organized as follows. In in Sec.~\ref{sec2}, we obtain the metric of EH BH surrounded by CDM halos, which has four parameters: mass $M$, charge $q$, QED parameter $a$, and CDM parameter $R$ and $\rho$. We discuss the influence of BH parameters on the optical properties of BH and plot the photon orbits of BH. In in Sec.~\ref{sec3}, we use the $(\varphi(b))$ diagram to analyze the imaging of BH accretion disks and plot the imaging of accretion disks seen by observers at different inclinations. Finally, we use the N-T model to plot the intensity imaging of BH accretion disks. We compare and demonstrate the effects of dark matter parameters on the imaging of BH accretion disks. Finally, we give a conclusion and outlook in Sec.~\ref{sec4}. To simplify the calculations, we use $c=G=M=1$ in this article.

	\section{Photon orbit}\label{sec2}
	\subsection{Metric of black hole}\label{sec2_1}
		The action for EH theory is written as \cite{EH2001}
	\begin{equation}\label{eq2_1}
		S={\frac{1}{4\pi}\int_{M^4}{d^4x\sqrt{-g}\left[ \frac{1}{4} R -\mathcal{L} \left( F,G \right) \right]}},
		\end{equation}
	\begin{equation}
		\mathcal{L} \left( F,G \right) =-F+\frac{a}{2}F^2+\frac{7a}{8}G^2.
	\end{equation}
	    Where $g$, $R$, are the determinant of the metric tensor, and Ricci scalar, respectively. $\mathcal{L} \left( F,G \right)$ is the NLED Lagrangian that depends on the electromagnetic invariants $F=\frac{1}{4}F_{\mu \nu}F^{\mu \nu}$ and $G=\frac{1}{4}{F_{\mu \nu}}^*F^{\mu \nu}$. $F_{\mu \nu}$ is the electromagnetic field strength tensor, and $^*F^{\mu \nu}$ is the dual form of $F_{\mu \nu}$.The non-zero components of the energy-momentum tensor of the EH black hole are \cite{EH2020,MSJ}
	\begin{equation}
		\begin{gathered}\label{teh}
	    	T_{t}^{t}(EH)=T_{r}^{r}(EH)=\frac{1}{4\pi}\left(-\frac{q^{2}}{2r^{4}}+\frac{aq^{4}}{8r^{8}}\right),\\
	    	T_{\theta}^{\theta}(EH)=T_{\phi}^{\phi}(EH)=\frac{1}{4\pi}\left(\frac{q^{2}}{2r^{4}}-\frac{3aq^{4}}{8r^{8}}\right).
	    \end{gathered}
    \end{equation}
        Where $q$ is electric charge of BH and $a$ is QED parameter. Now, we have the non-zero components of the energy-momentum tensor for CDM halo as \cite{CDMT}
    \begin{equation}\label{tcdm}
    	\begin{gathered}
    	T_{t}^{t}(CDM)=T_{r}^{r}(CDM)=\frac{1}{8 \pi}g(r)(\frac{1}{r}\frac{g^{'}(r)}{g(r)}+\frac{1}{r^{2}})-\frac{1}{r^{2}},\\
    	\begin{split}
    		T_{\theta}^{\theta}(CDM)=T_{\phi}^{\phi}(CDM)&=\frac{1}{16 \pi}g(r)[\frac{g^{\prime\prime}(r)g(r)-g^{\prime2}(r)}{g^{2}(r)}\\ &+\frac{1}{2}\frac{g^{\prime2}(r)}{g^{2}(r)}
    		+\frac{1}{r}(\frac{g^{\prime}(r)}{g(r)}+\frac{g^{\prime}(r)}{g(r)})+\frac{g^{\prime}(r)g^{\prime}(r)}{2g(r)g(r)}],
    	\end{split}
        \end{gathered}
    \end{equation}
        and
    \begin{equation}
    	g(r)=(1+\frac{r}{R})^{-\frac{8\pi \rho R^{3}}{r}},
    \end{equation}
        where $R$ is feature radius, $\rho$ is density of the CDM halo collapse. We substitute \ref{teh} and \ref{tcdm} into the Einstein field equations
    \begin{equation}
    	R^{\nu}{}_{\mu}-\frac{1}{2}\delta^{\nu}{}_{\mu}R=8 \pi T^{\nu}{}_{\mu}=8 \pi [T^{\nu}{}_{\mu}(EH)+T^{\nu}{}_{\mu}(CDM)].
    \end{equation}
		Ultimately, we obtain the line element of EH BH in the CDM halo as
	\begin{equation}\label{eq2_2}
		\begin{aligned}
			ds^2&=-f\left(r\right)dt^2+\frac{1}{f\left(r\right)}dr^2+r^{2}d\theta^{2}+r^{2}\sin^{2}{\theta}d\phi^{2},\\
			&f\left(r\right)=\left(1+\frac{r}{R}\right)^{-\frac{8\pi\rho R^{3}}{r}}-\frac{2M}{r}+\frac{q^2}{r^2}-\frac{aq^4}{20r^6},
		\end{aligned}
	\end{equation}
		where $M$ is mass of BH.
	\subsection{Geodesic equation}\label{sec2_2}
		For static spherically symmetric space-time, line element is
	\begin{equation}\label{eq2_3}
		ds^{2}=g_{\mu\nu}dx^{\mu}dx^{\nu}=-f\left(r\right)dt^{2}+\frac{1}{f\left(r\right)}dr^{2}+r^{2}d\theta^{2}+r^{2}\sin^{2}{\theta}d\phi^{2}.
	\end{equation}
	A particle's Lagrangian is
	\begin{equation}\label{eq2_4}
		\mathcal{L}=\frac{1}{2}g_{\mu\nu}\dot{x}^{\mu}\dot{x}^{\nu}=\frac{1}{2}\left(-f\left(r\right)\dot{t}^2+\frac{1}{f\left(r\right)}\dot{r}^2+r^2\dot{\theta}^2+r^{2}\sin^{2}{\theta}\dot{\phi}^{2}\right),
	\end{equation}
		where $\dot{x}^{\mu}=\frac{dx^{\mu}}{d\lambda}$. For photon $\lambda$ is affine parameter and for time-like particle, $\lambda$ is proper time $\tau$. There are two Killing vector fields in static spherically symmetric space-time $\frac{\partial}{\partial t}$ and $\frac{\partial}{\partial \phi}$, so particle has its two conserved quantities
	\begin{equation}\label{eq2_5}
		E=-\frac{\partial \mathcal{L}}{\partial \dot{t}}=f\left(r\right)\dot{t},
		\end{equation}
		\begin{equation}\label{eq2_6}
		L=\frac{\partial \mathcal{L}}{\partial \dot{\phi}}=r^{2}\dot{\phi},
		\end{equation}
		where $E$ is particle's energy and $L$ is angular momentum. Please notice that we have chosen $\theta=\frac{\pi}{2}$, which means that particle always moves on equatorial plane. Impact parameter $b$ is defined as
	\begin{equation}\label{eq2_7}
		b:=\frac{|L|}{E}.
	\end{equation}
		For photon, $b$ is the distance from the asymptotic line of the light orbit to the center of the black hole at infinity. Photon's motion meets $\mathcal{L}=0$, use Eq.~\ref{eq2_5}, Eq.~\ref{eq2_6} and reparameterize affiine parameter $\lambda'=L\lambda$, the equation of motion is written as
	\begin{equation}\label{eq2_8}
		\dot{t}=\frac{1}{bf\left(r\right)},
	\end{equation}
	\begin{equation}\label{eq2_9}
		\dot{\phi}=\pm\frac{1}{r^2},
	\end{equation}
	\begin{equation}\label{eq2_10}
		\dot{r}^{2}=\frac{1}{b^2}-\frac{f\left(r\right)}{r^2},
	\end{equation}
		where $\pm$ indicates two different directions of photon's rotation and they have no essential distinctions. Effective potential $V_{eff}$ for photon is defined as
	\begin{equation}\label{eq2_11}
		V_{eff}:=\frac{f\left(r\right)}{r^2}.
	\end{equation}
		According to analytical mechanics, photon sphere should meet $\dot{r}=0,V_{eff}'=0$ (we always mark $\frac{dA}{dr}$ as $A'$ for any functions $A\left(r\right)$), then radius of photon sphere $r_{ph}$ satisfies
	\begin{equation}\label{eq2_12}
		V_{eff}'\big|_{r=r_{ph}}=0,
	\end{equation}
	\begin{equation}\label{eq2_13}
		b_{c}=\frac{r_{ph}}{\sqrt{f\left(r_{ph}\right)}}.
	\end{equation}
		Here $b_{c}$ is impact parameter corresponding to the photon moving in photon sphere. Now it's necessary to introduce a crucial concept, innermost stable circular orbit (ISCO) for time-like particle. Similar to process we have done for null geodesic, equation of motion for physical particle ($\mathcal{L}=-\frac{1}{2}$) is
	\begin{equation}\label{eq2_14}
		\dot{r}^{2}=\frac{1}{b^2}-\frac{f\left(r\right)}{r^2}-\frac{f\left(r\right)}{L^2},
	\end{equation}
		Particle's energy $E$, angular momentum and impact parameter still take Eq.~\ref{eq2_5}, Eq.~\ref{eq2_6} and Eq.~\ref{eq2_7}. Here $\dot{x}^{\mu}=\frac{dx^{\mu}}{d\tau'}$ and $\tau'=L\tau$, where $\tau$ is proper time for time-like particle. The radius of ISCO $r_{isco}$, can be calculated by
	\begin{equation}\label{eq2_15}
		\dot{r}^{2}=\left(\dot{r}^{2}\right)'=\left(\dot{r}^{2}\right)''=0.
	\end{equation}
		$r_{isco}$ is a critical value when we study the image of thin accretion disk in next section. It actually represent the inner edge of accretion disk, because circular orbits within this range will be unstable, such that the particle will fall into the black hole or escape to flee farther when it is perturbed.
	
	\subsection{Influence of parameters on black hole}\label{sec2_3}
		Before we discuss the photon orbit, we first consider the influences of four parameters $q$, $a$, $R$ and $\rho$ on BH.
	\begin{table}[htbp]
		\centering
		\includegraphics[width=0.6\textwidth]{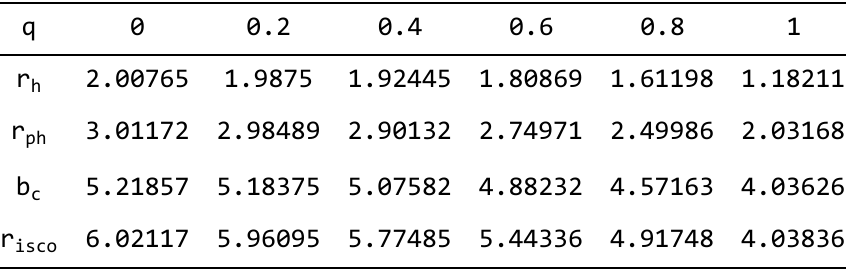}
		\caption{Variations of $r_{h}$, $r_{ph}$, $b_{c}$ and $r_{isco}$ with respect to $q$. We set $a=1$, $R=0.1$ and $\rho=0.1$.}\label{tab_q}
	\end{table}
	\begin{table}[htbp]
		\centering
		\includegraphics[width=0.6\textwidth]{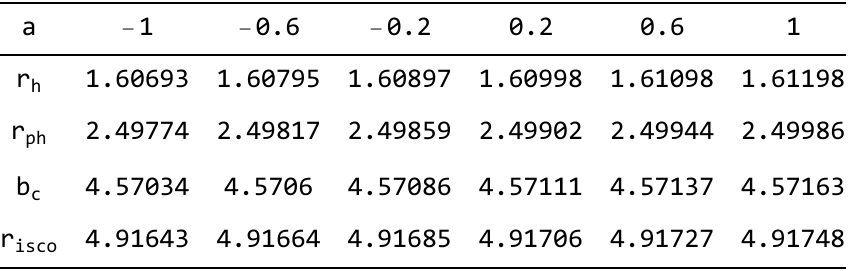}
		\caption{Variations of $r_{h}$, $r_{ph}$, $b_{c}$ and $r_{isco}$ with respect to $a$. We set $q=0.8$, $R=0.1$ and $\rho=0.1$.}\label{tab_a}
	\end{table}
	
		Table.~\ref{tab_q}, Table.~\ref{tab_a} give the influences of $q$ and $a$ on radius of event horizon $r_{h}$, which meets $f\left(r_{h}\right)=0$, $r_{ph}$, $b_{c}$ and $r_{isco}$. As seen in table, $r_{h}$, $r_{ph}$, $b_{c}$ and $r_{isco}$ will increase with the increase of $a$ and decrease of $q$, but the influence of $a$ is so small. So we fix $a=1$ and set $q=0,0.5,0.8$ to investigate the influences of $R$ and $\rho$. Table.~\ref{tab_rhoR} shows the influences of $R$ and $\rho$ on $r_{h}$, $r_{ph}$, $b_{c}$ and $r_{isco}$.
	\begin{table}[htbp]
		\centering
		\includegraphics[width=0.96\textwidth]{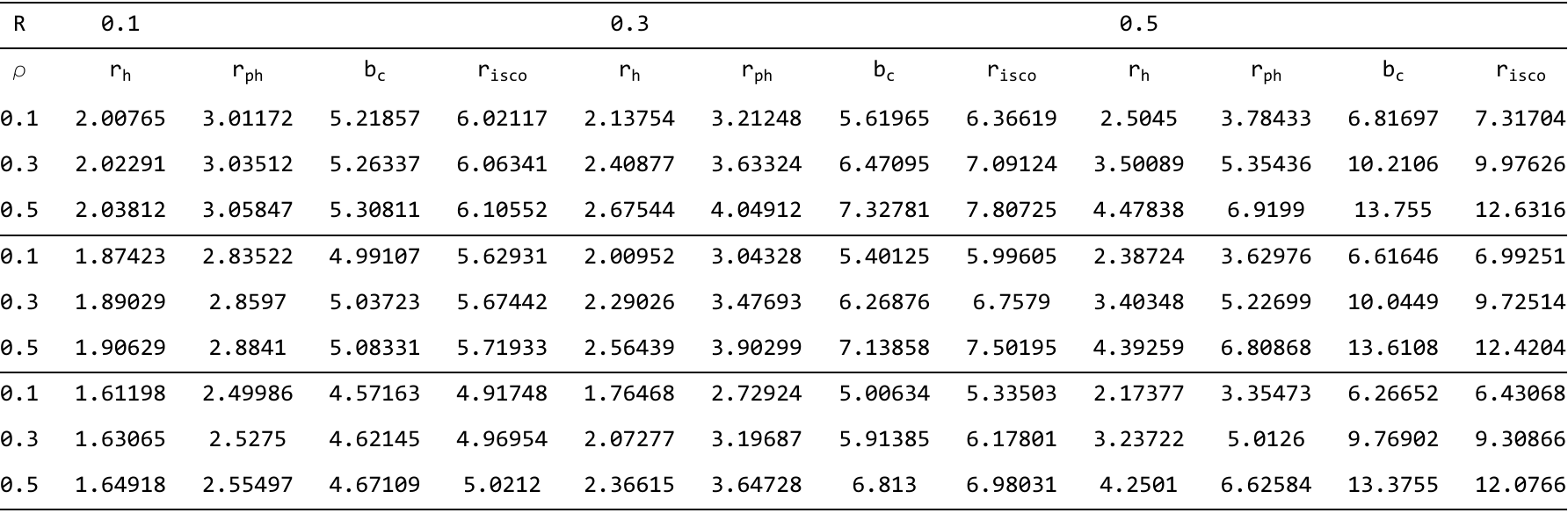}
		\caption{Variations of $r_{h}$, $r_{ph}$, $b_{c}$ and $r_{isco}$ with respect to $R$ and $\rho$. We set $a=1$, and values of $q$ in three sections of from top to bottom are $0$, $0.5$ and $0.8$ respectively.}\label{tab_rhoR}	\end{table}
		
		\noindent It is clear in table that effect of CDM halo is so evident, especially change of $R$. The increase of $R$ and $\rho$ will distinctly increase the values of $r_{h}$, $r_{ph}$, $b_{c}$ and $r_{isco}$. In our future work, we will choose two sets of parameters with transparent differences to research the photon orbit.
		
\begin{figure}[htbp]
		\centering
		\includegraphics[width=1\textwidth]{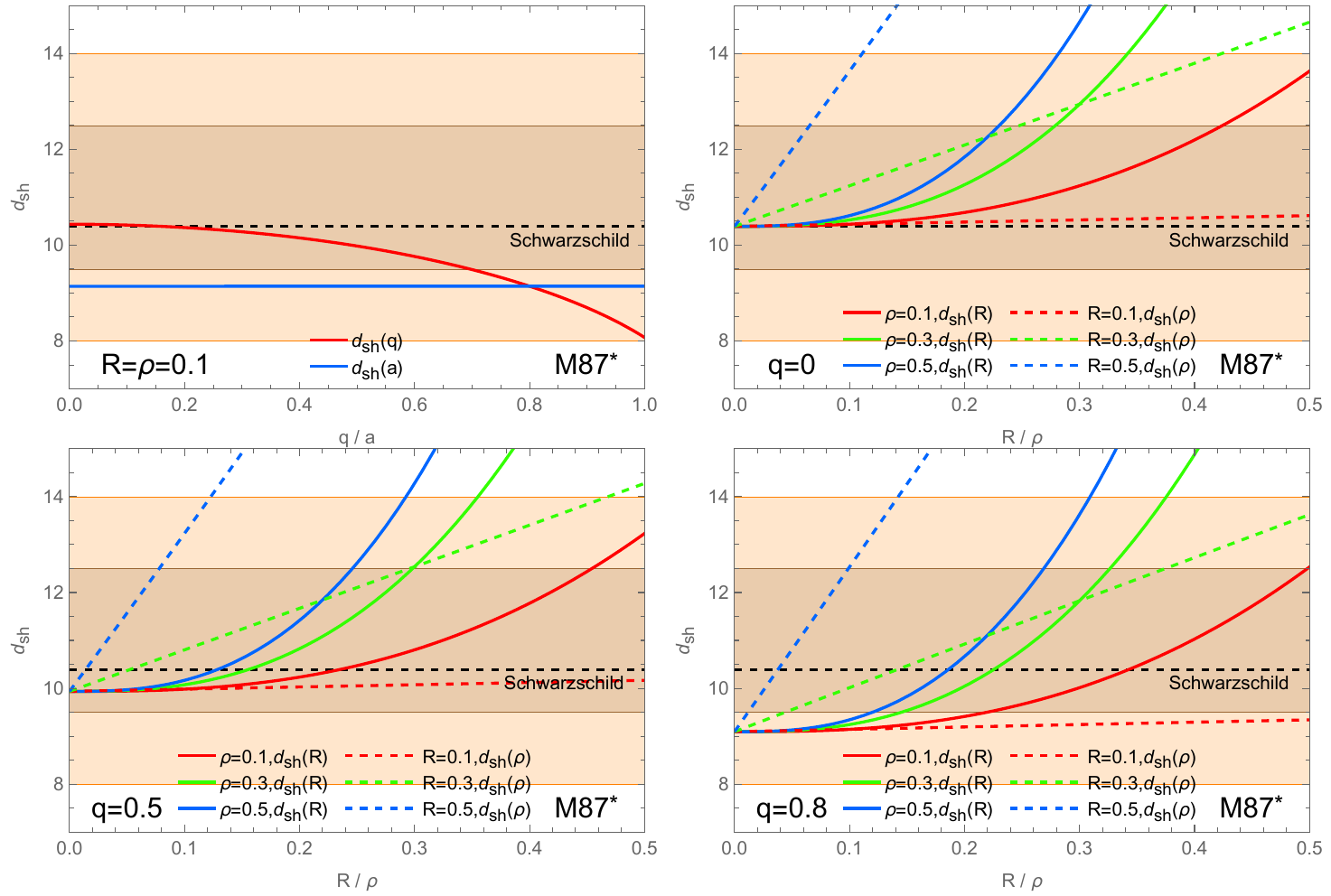}
		\caption{Shadow diameter of the CDM-EH BH as a function of the deformation parameter, The orange and
			brown shaded regions represent the regions of $1\sigma$ and $2\sigma$ confidence intervals, respectively, with
			respect to the M87* observations}\label{fig_M87}
	\end{figure}

		Following the variations of parameters outlined above, we plotted in Fig.~\ref{fig_M87} the function describing the BH's shadow diameter as a function of these parameters. We then compared these function plots with the M87* BH shadow data obtained from the EHT, thereby constraining the parameters of our BH model. As is widely acknowledged, the angular size of the M87* shadow is denoted by $\delta~=~(42\pm3)~\mu\text{as}$, its distance by $D=16.8_{-0.7}^{+0.8}$, and its mass by $M=\begin{matrix}(6.5\pm0.9)\times10^9M_\odot\end{matrix}$. By utilizing this data, the shadow diameter $d_{M87^{*}}=\frac{D\delta}{M}\simeq11.0\pm1.5$ of the black hole can be computed \cite{DM871,DM872}.
	\subsection{Photon trajectory}\label{sec2_4}
		We use the data in Table.~\ref{tab_rhoR} to plot the effective potential for photon.
	\begin{figure}[htbp]
			\centering
			\includegraphics[width=1\textwidth]{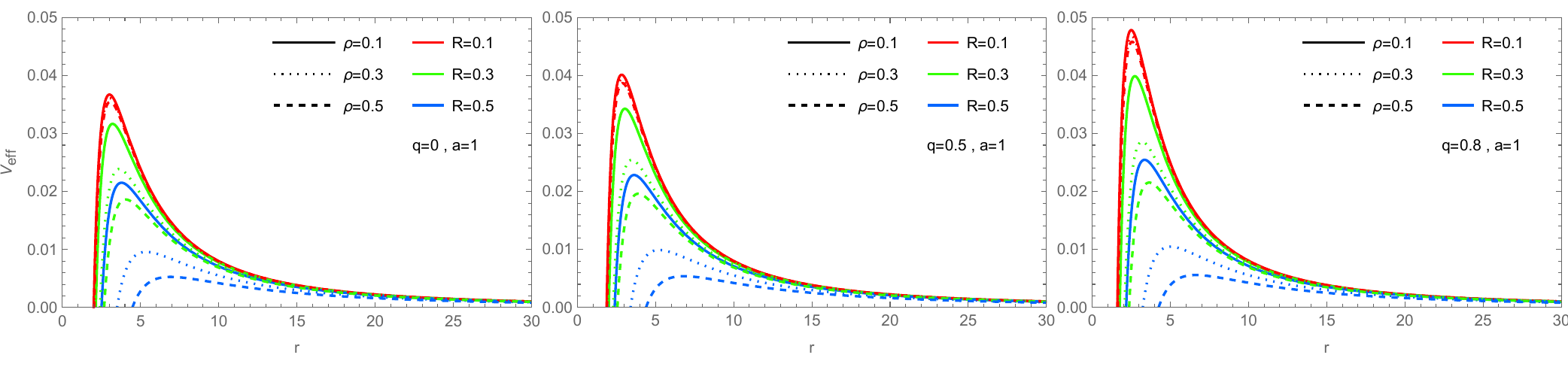}
			\caption{Variations of $V_{eff}$ with respect to $r$.}\label{fig_Veff}
	\end{figure}
		Figure.~\ref{fig_Veff} shows the influences of $q$, $R$ and $\rho$ on the variations of $V_{eff}$ when $a=1$. One could find in figure that $V_{eff}$ increases with the increase of $q$. Of note is that the increase of $R$ and $\rho$ will restrain $V_{eff}$, and these inhibitory effects will become obvious when $R$ and $\rho$ are bigger. Under the restriction of $R$ and $\rho$, the two blue lines at the bottom of figure have hardly exhibited the impact of $q$. Meanwhile, the variations of $r_h$ and $r_{ph}$ with respect to parameters in Fig.~\ref{fig_Veff} agree with the data in Table.~\ref{fig_Veff}.
		
		Now we calculate the photon orbit in detail. Use Eq.~\ref{eq2_9}, Eq.~\ref{eq2_10} and make a transform $u=\frac{1}{r}$ we obtain the equation of motion for photon
	\begin{equation}\label{eq2_16}
		G\left(u\right):=\left(\frac{du}{d\phi}\right)^{2}=\frac{1}{b^{2}}-u^{2}f\left(\frac{1}{u}\right).
	\end{equation}
		We take $q=0.5, a=1$ for an example to show how $R$ and $\rho$ affect $G\left(u\right)$.
	\begin{figure}[htbp]
		\centering
		\includegraphics[width=1\textwidth]{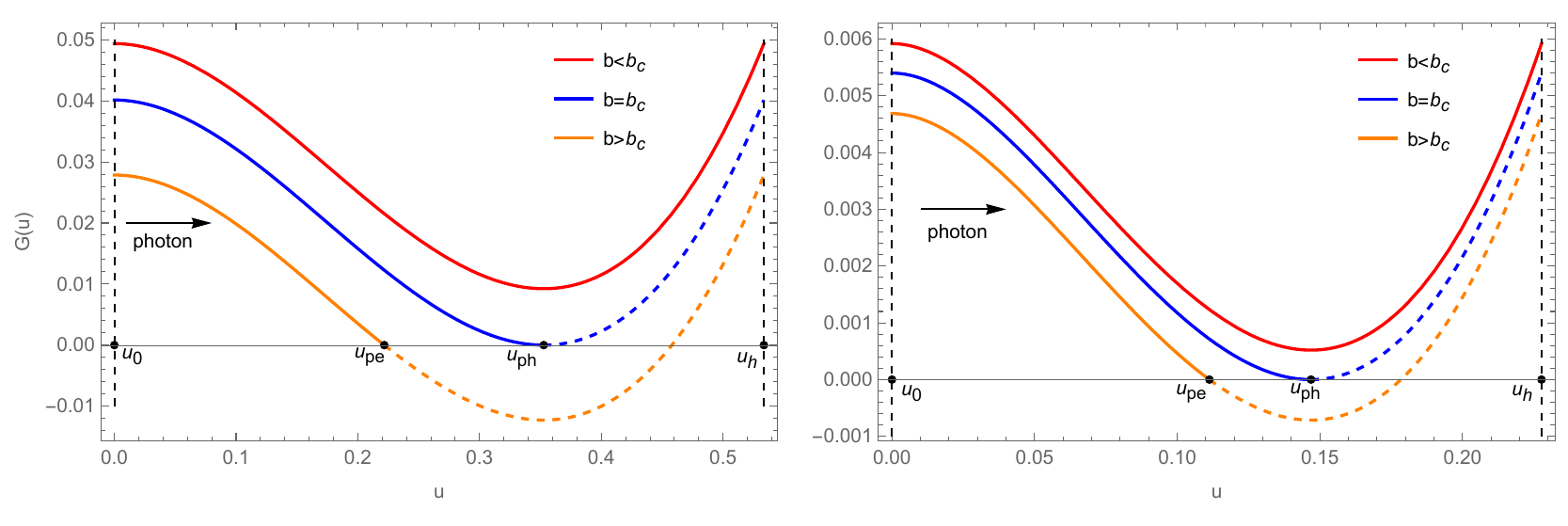}
		\caption{Functions $G\left(u\right)$ of $u$. We set $q=0.5, a=1$. The left side corresponds to $R=\rho=0.1$, and the right side corresponds to $R=\rho=0.5$. $u_{0}=0$, $u_{ph}=\frac{1}{r_{ph}}$, $u_{h}=\frac{1}{r_{h}}$ and $u_{pe}$ is the minimum positive root of $G\left(u\right)$ when $b>b_{c}$.}\label{fig_Gu}
	\end{figure}
		One could understand photon's orbit via Fig.~\ref{fig_Gu}:
		
		For $b>b_c$, there is a minimum positive root $u_{pe}$ of $G\left(u\right)$. The photon coming from infinity will reach its perihelion $r_{pe}=\frac{1}{u_{pe}}$ and then return to infinity along a path with the same shape as the incoming trajectory, Eq.~\ref{eq2_16} gives the total change of azimuth angle $\varphi$ (We always use $\phi$ and $\varphi$ to indicate the space-time coordinate and the change of $\phi$ in photon's motion respectively):
	\begin{equation}\label{eq2_17}
		\varphi=2\int_{0}^{u_{pe}}\frac{1}{\sqrt{G\left(u\right)}}du.
	\end{equation}
		For photon corresponding to $b<b_c$, the photon will continuously approach the black hole until they fall into the horizon. The total change of azimuth angle $\varphi$ for this kind of photons is
	\begin{equation}\label{eq2_18}
			\varphi=\int_{0}^{u_{h}}\frac{1}{\sqrt{G\left(u\right)}}du.
	\end{equation}
		For photon corresponding to $b=b_c$, it will arrive at $u_{ph}$ and then perpetually undergo circular motion.
		
\begin{figure}[htbp]
		\centering
		\includegraphics[width=1\textwidth]{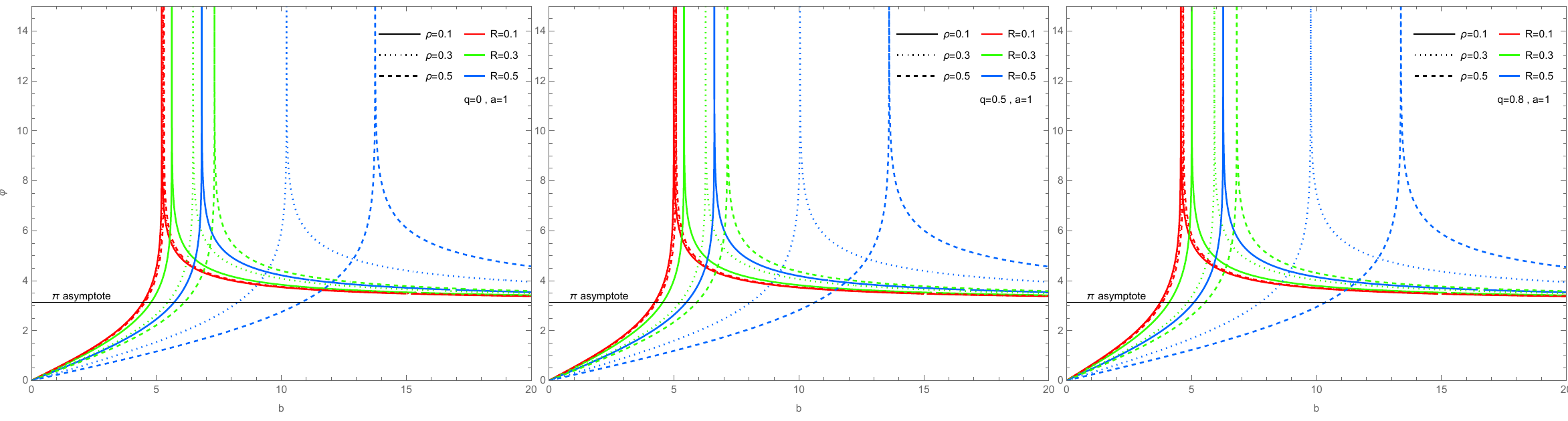}
		\caption{$\varphi\left(b\right)$ for different parameters.}\label{fig_phib}
	\end{figure}

		Figure.~\ref{fig_phib} gives the $\varphi\left(b\right)$ for different parameters.
		As shown in the figure, $\lim_{b\rightarrow\infty}\varphi=\pi$, which corresponds with flat space-time at infinity. The region of $b$ will be narrow with the increase of $\varphi$, which will shrink the photon's image, making the higher-order images of accretion disk hard to distinguish.
		
		The increase of $q$ will decrease the value of $b_{c}$. And as $R$ and $\rho$ rise, the entire image of $\varphi\left(b\right)$ will move towards the upper right direction, especially when $R$ and $\rho$ become more bigger. All of this will make a great contribution to BH's optical properties.
		
\begin{figure}[htbp]
		\centering
		\includegraphics[width=1\textwidth]{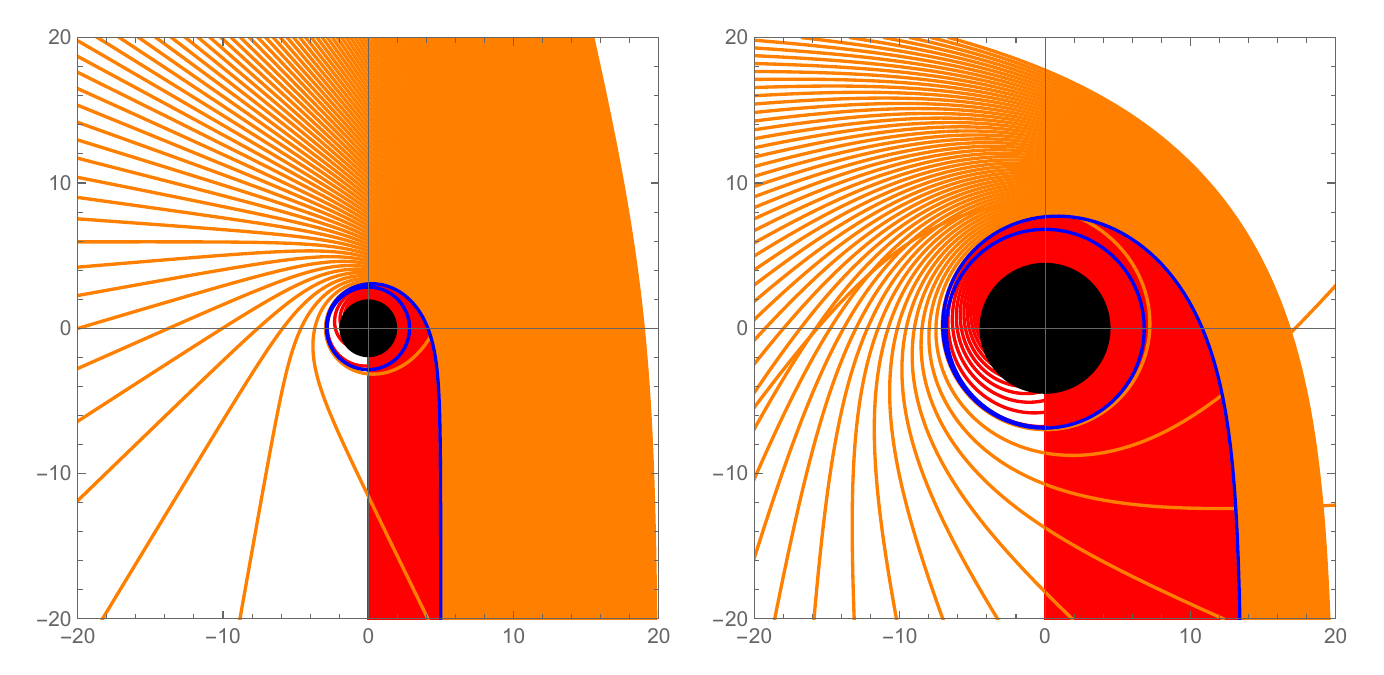}
		\caption{Photon trajectories for different parameters. We set $q=0.5$ and $a=1$. The left side corresponds to $R=\rho=0.1$ and the right side corresponds to $R=\rho=0.5$. The red, blue and orange curves represent the photons which meet $b<b_{c}$, $b=b_{c}$ and $b>b_{c}$ respectively.}\label{fig_photonorbit}
	\end{figure}

		At last, we plot the photon trajectories for different parameters as Fig.~\ref{fig_photonorbit}.
	\newpage
	\section{Image of thin accretion disk}\label{sec3}
	\subsection{Observation coordinate system}\label{sec3_1}
	\begin{figure}
		\centering
		\includegraphics[width=1\textwidth]{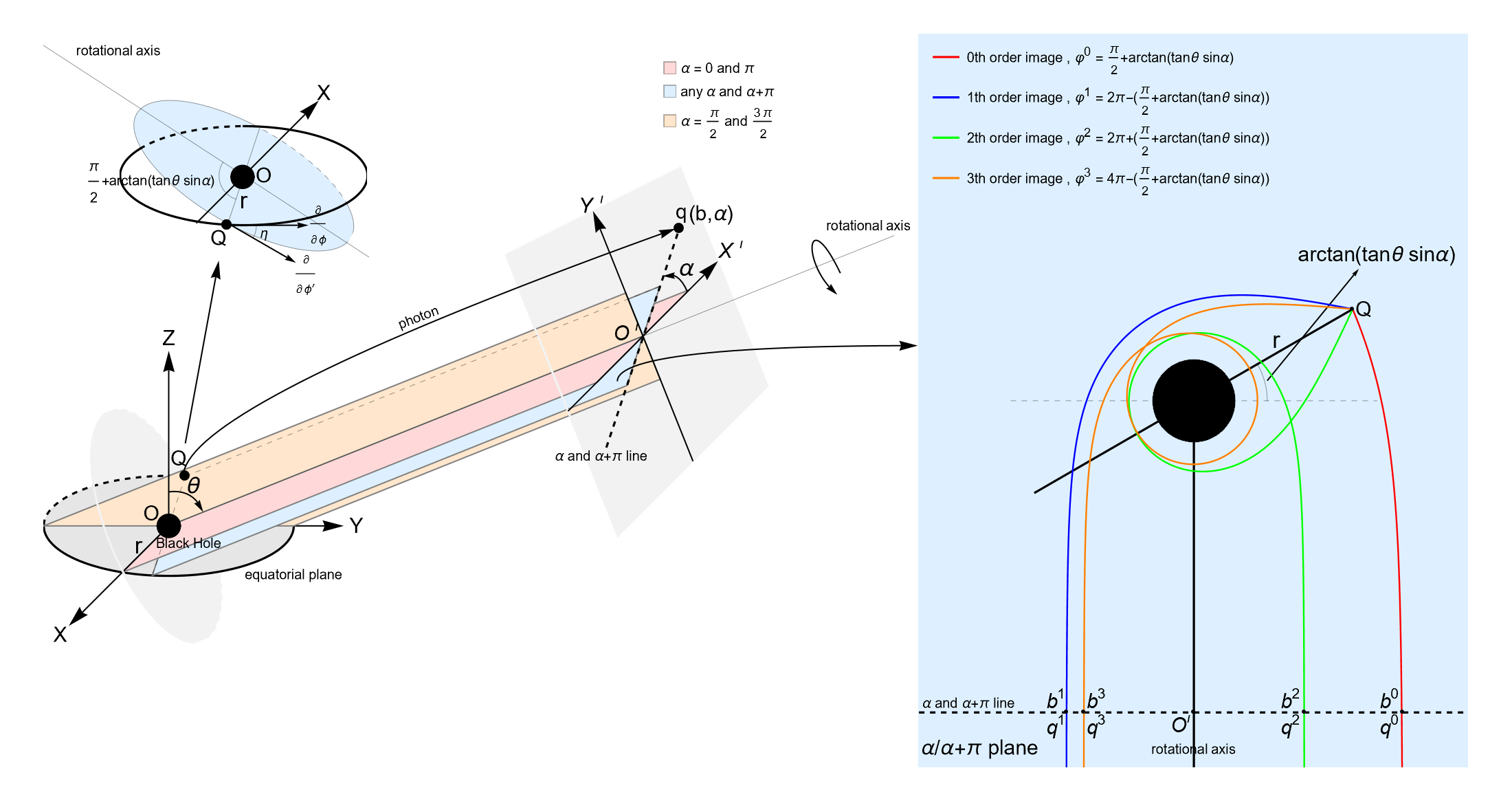}
		\caption{Coordinate system.}\label{fig_coordinate}
	\end{figure}
		To investigate the image of thin accretion disk, the observation coordinate system is shown as Fig.~\ref{fig_coordinate}. The observer locates at $\left(\infty,\theta,0\right)$ in the black hole's spherical coordinate system $\left(r,\theta,\phi\right)$, which takes center of BH as $r=0$. Consider in the observer coordinate system $O'X'Y'$, a photon departing from the point $q\left(b,\alpha\right)$ in the vertical direction, which means that $b$ is precisely photon's impact parameter. This photon arrives at a point $Q\left(r,\frac{\pi}{2},\phi\right)$ in accretion disk. Because of the reversibility of light path, the photon from $Q\left(r,\frac{\pi}{2},\phi\right)$ will finally reach the its image point $q\left(b,\alpha\right)$.
		\par
		If we fix $r$, we will derive the image of equal-$r$ orbit. As shown in the left side of Fig.~\ref{fig_coordinate}, every $\alpha/\alpha+\pi$ plane and equal-$r$ orbit in equatorial plane have two intersection points with the difference of azimuth angel $\phi$ being $\pi$. We set $\alpha=0$ and $\phi=0$ for $X'$-axis and $X$-axis respectively. According to geometry, the angle $\varphi$ between rotational axis and $OQ$ is
	\begin{equation}\label{eq3_1}
		\varphi=\frac{\pi}{2}+\arctan\left(\tan\theta\sin\alpha\right).
	\end{equation}
		When $b$ gets closer to $b_{c}$, the light will bend more severely. So a source point $Q$ could have many image points $q$. We label these image points according to their $\varphi$ from small to large as $q^{n}$ ($n\in \mathbb{N}$), which represents $n^{th}$-order image respectively. As shown in the right side of Fig.~\ref{fig_coordinate}, all the even-order images of $Q$ are in the same side ($\alpha$) as $Q$. In contrast, all the odd-order images of $Q$ will appear in the opposite side ($\alpha+\pi$). We mark photon's changes of $\phi$ resulting in the $n^{\rm{th}}$-order image as $\varphi^{n}$: 
	\begin{equation}\label{eq3_2}
		\varphi ^n=
		\begin{cases}
		\frac{n}{2} 2\pi +\left( -1 \right) ^n\left[ \frac{\pi}{2}+\mathrm{arc}\tan \left( \tan \theta \sin \alpha \right) \right]&,\mathrm{when}~n~\mathrm{is~ even},\\
		\frac{n+1}{2} 2\pi +\left( -1 \right) ^n\left[ \frac{\pi}{2}+\mathrm{arc}\tan \left( \tan \theta \sin \alpha \right) \right]&,{\mathrm{when}~n~\mathrm{is~ odd}}.
		\end{cases}		
	\end{equation}
	 	Substituting these $\varphi^{n}$ into Eq.~\ref{eq2_16} one can get the their corresponding impact parameters $b^{n}$. The image point of source point $Q$ in observer coordinate system $O'X'Y'$ can be expressed as $q^{n}\left(b^{n},\alpha\right)$ for even number $n$ and $q^{n}\left(b^{n},\alpha+\pi\right)$ for odd number $n$.
	\subsection{Image of equal-$r$ orbit on thin accretion disk}\label{sec3_2}
	\subsubsection{Intersection condition of photon's orbit and equal-$r$ orbit}\label{sec3_2_1}
\begin{figure}[h]
			\centering
			\includegraphics[width=1\textwidth]{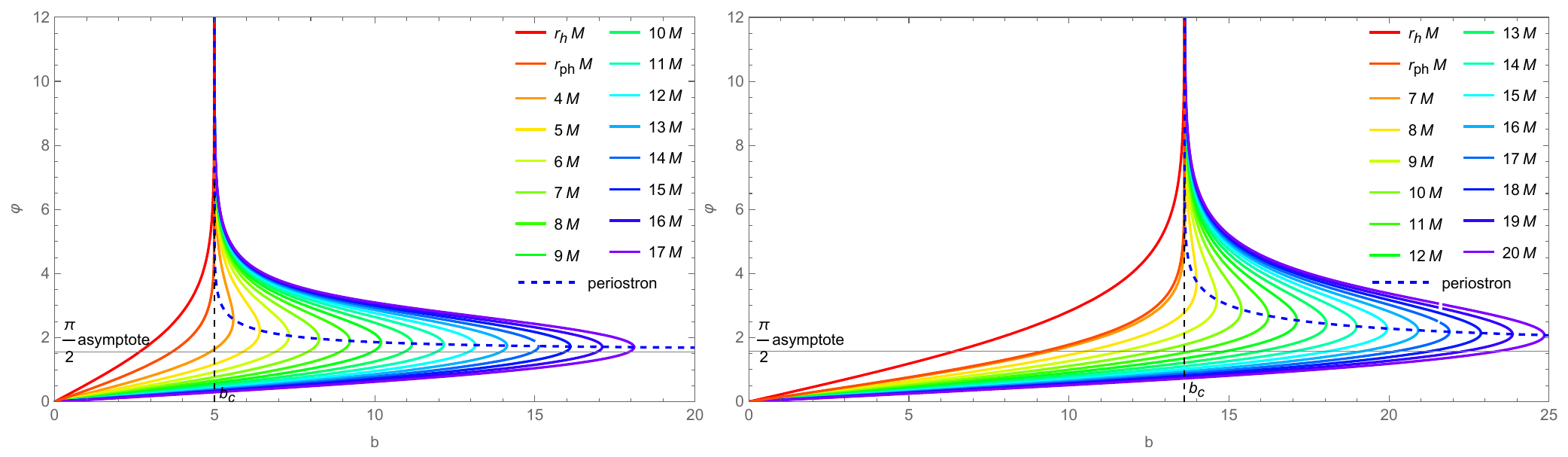}
			\caption{Deflection angle $\varphi\left(b\right)$ corresponding to intersections as a function of $b$ for different $r$. We set $q=0.5, a=1,R=\rho=0.1$ for the left side and $q=0.5, a=1,R=\rho=0.5$ for the right side.}\label{fig_rphib}
		\end{figure}

		For photons coming form infinity with different values of $b$ on their trajectory plane, they will have different intersections with equal-$r$ orbit.
		Fig.~\ref{fig_rphib} gives figure of $\varphi\left(b\right)$. We denote the blue dashed line as $\varphi_{blue}(b)$. Using the blue dashed line as the demarcation, we designate the colored curves below the demarcation line as $\varphi _{color-down}(b)$ and those above it as $\varphi _{color-up}(b)$. Thus, we have:
		\begin{equation}
			\varphi_{blue} \left( b \right) =\int_0^{u_{pe}}{\frac{1}{\sqrt{\frac{1}{b^2}-f\left( \frac{1}{u} \right)}}}du
		\end{equation}
		
		\begin{equation}\label{cd}
			\varphi _{color-down}\left( b \right) =\int_0^{u_{r}}{\frac{1}{\sqrt{\frac{1}{b^2}-f\left( \frac{1}{u} \right)}}}du
		\end{equation}
		
		\begin{equation}
			\begin{split}
				\varphi _{color-up}\left( b \right) &=2\int_0^{u_{pe}}{\frac{1}{\sqrt{\frac{1}{b^2}-f\left( \frac{1}{u} \right)}}}du-\int_0^{u_r}{\frac{1}{\sqrt{\frac{1}{b^2}-f\left( \frac{1}{u} \right)}}}du\\
				&=2\varphi_{blue}(b)-\varphi _{color-down}(b)
			\end{split}
		\end{equation}
		
		In the figures, every colored line represents a equal-$r$ orbit, point $ \left(b, \varphi \right)$ in colored line indicates that the deflection angel when photon that takes $b$ as its impact parameter arrives at equal-$r$ is $\varphi$. Blue dashed line will intersect with colored lines at their peaks, meaning that the point $\left(b, \varphi \right)$ on blue dashed line represents that the deflection angel when photon that takes $b$ as its impact parameter arrives at its perihelion $r_{pe}$. And it is clear that blue dashed lines will take $\varphi=\frac{\pi}{2}$ as its asymptotic line. It correspond to that the photon corresponding to $b=\infty$ will move along a straight line, which and circle $r=\infty$ are tangent at point $\varphi=\frac{\pi}{2}$.

		By comparing two figures with different parameters, one could find on the left side, the colored lines are stretched drastically when $r>r_{ph}$ and the right side is opposite. This contrast will cause that maximum of image for equal-$r$ will tend to $\alpha=0,\pi$ more fast, which leads to different images of a fixed equal-$r$ orbit for different parameters.
		
		Now, we fix $b$ to investigate the intersection condition of photon's orbit and equal-$r$ orbit according to Fig.~\ref{fig_rphib}. Obviously the photon corresponding to $b<b_{c}$ will intersect with equal-$r$ orbit for only once without perihelion. It could be divided into three cases if $b\geq b_{c}$:
		
		\noindent(i) $r_{h}<r<r_{ph}$. When $b\geq b_{c}$, the photon orbit has no intersection with equal-$r$ orbit.
		
		\noindent(ii) $r=r_{ph}$. When $b=b_{c}$, the photon orbit will have infinite intersections because it will undergo circular motion $r=r_{be}=r_{ph}$ forever. When $b>b_{c}$, the photon orbit still has no intersections with equal-$r$ orbit;
		
		\noindent(iii) $r>r_{ph}$, there is a critical value $b_{pe}$ , whose corresponding $r_{pe}$ is $r$. When $b_c<b<b_{pe}$, the photon has two intersections with equal-$r$ orbit, whose $\varphi\left(b\right)$ are marked as $\varphi_{1}\left(b\right)$ and $\varphi_{2}\left(b\right)$ by us. When $b=b_{pe}$, photon orbit and equal-$r$ orbit intersect at one point. When $b>b_{pe}$, photon orbit has no intersections with equal-$r$ orbit. For $b<b_{pe}$, when $b$ gets closer to $b_c$, $\varphi_{2}\left(b\right)$ will get bigger. And when $r\rightarrow\infty$, $\varphi\left(b_{pe}\right)\rightarrow\frac{\pi}{2}$, $\varphi\left(b\right)\rightarrow \arcsin\left(\frac{b}{r}\right)$.
		
		\noindent As a summary, we give Table.~\ref{tab_intersection} to show the number of intersections for different $b$ and equal-$r$ orbit.
	\begin{table}[h]
		\centering
		\includegraphics[width=0.5\textwidth]{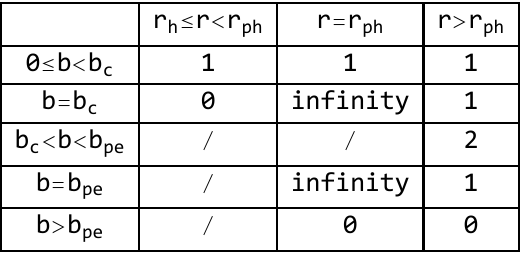}
		\caption{Number of intersections between photon orbit and equal-$r$ orbit for different $b$ and $r$.}\label{tab_intersection}
	\end{table}

	\subsubsection{Image of equal-$r$ orbit}\label{sec3_2_2}
	
		To investigate the image of equal-$r$ orbit, we plot inverse function $b\left(\varphi\right)$ and $\alpha\left(\varphi\right)$ as Fig.~\ref{fig_1bphi}. We take $\theta=0$/$\frac{\pi}{6}$/$\frac{\pi}{3}$/$\frac{4\pi}{9}$ for an example and give these following analyses:
		\begin{figure}[htbp]
			\centering
			\includegraphics[width=1\textwidth]{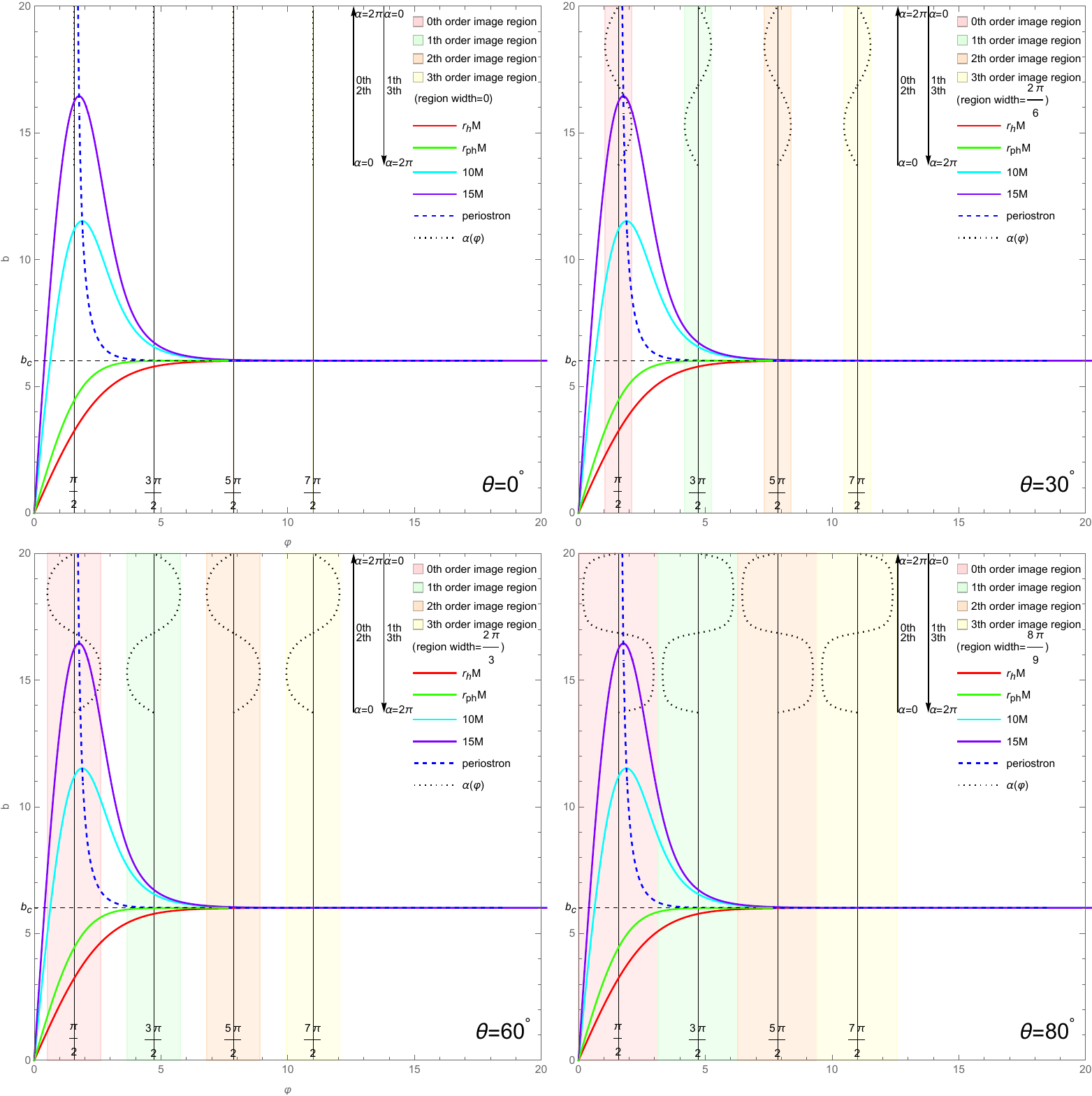}
			\caption{Figure of $b\left(\varphi\right)$. S-shaped curves are  inverse functions of Eq.~\ref{eq3_2} corresponding to respective images. For every even-order image, $\alpha$ takes $0$ to $2\pi$ form bottom to top. And odd-order image is opposite. We set $q=0.5$, $a=1$ and $R=\rho=0.1$.}\label{fig_1bphi}
		\end{figure}
		\newpage
		\begin{itemize}
		\item$\theta=0$\
		\begin{itemize}
		\item[(a)] For any $\alpha$, $\varphi$ will be always $\frac{\pi}{2}$, which means the image of equal-$r$ orbit is a perfect circle. It is reasonable in physics due to the symmetry;
		\item[(b)] With the increase of images' order $n$, $\Delta b$ caused by two different values of $r$ will drastically decrease, which makes higher-order images hard to distinguish, especially for $n\geq2$.
		\item[(c)] The maximum $b_{\rm{max}}$ of colored lines (equal-$r$ orbit) will also decrease as $n$ increases, and $b_{\rm{max}}$ will get close to $b_{c}$ so quickly when $n\geq 2$. For any $r$, $0^{\rm{th}}$ image of equal-$r$ orbit will never get $b_{pe}$ unless $r=\infty$.
		\end{itemize}
		\item $\theta=\frac{\pi}{6}/\frac{\pi}{3}/\frac{4\pi}{9}$
		\begin{itemize}
		\item [(a)] Due to nonzero inclination angle, the interval of photon's deflection angle $\varphi$ has a width, which leads to that its image is no longer a circle. But it can be seen in the figure that image is symmetric about Y$'$-axis resulted by S-shaped curves. However it is expectable that when $\theta$ goes to $\frac{\pi}{2}$, the images of equal-$r$ orbit will return to a perfect circle, we will give a detailed discussion at the end of this section;
		\item [(b)] Similar to the case for $\theta=0$, the $n^{\rm{th}}$ images are hardly distinguishable for $n\geq 2$. But when $n\geq 1$, compared to $\theta=0$, for images of each order, $\Delta b$ is no longer a constant. And one could see in the figures that $\Delta b_{\rm{max}}$ and $\Delta b_{\rm{min}}$ will increase and decrease respectively when $n$ becomes bigger. For a fixed $n$-order image, as $\theta$ increases, the images for $\alpha\in[0,\pi]$ are hard to differentiate and the images for $\alpha\in[\pi,2\pi]$ are more easy to distinguish. In contrast, when $n=0$, images for any $\alpha\in[0,2\pi]$ will become hard to distinguish with the increase of $\theta$, especially for region $\alpha\in[\pi,2\pi]$.
		\item [(c)] The same as our discussion above, the behaviour of $b_{\rm{max}}$ is similar with $\Delta b_{\rm{max}}$. Due to the nonzero width of $\varphi$, some images of equal-$r$ orbit have these corresponding perihelions $b_{pe}$. It should be pointed out that once the image acquires its $b_{pe}$, $b_{pe}$ will exactly become $b_{\rm{max}}$. Specially, for $0^{\rm{th}}$-order image, when $r\rightarrow\infty$, $b_{\rm{max}}=b_{pe}$ will get close to X$'$-axis, which is related to an ellipse image resulted by the flat space-time at infinity.

		And for any $r<r_{ph}$, image for equal-$r$ orbit will achieve their $b_{\rm{min}}$ and $b_{\rm{max}}$ when $\alpha=\frac{3\pi}{2}$ and $\alpha=\frac{\pi}{2}$ respectively.
		\end{itemize}
		\end{itemize}
		\par
\begin{figure}[htbp]
		\centering
		\includegraphics[width=1\textwidth]{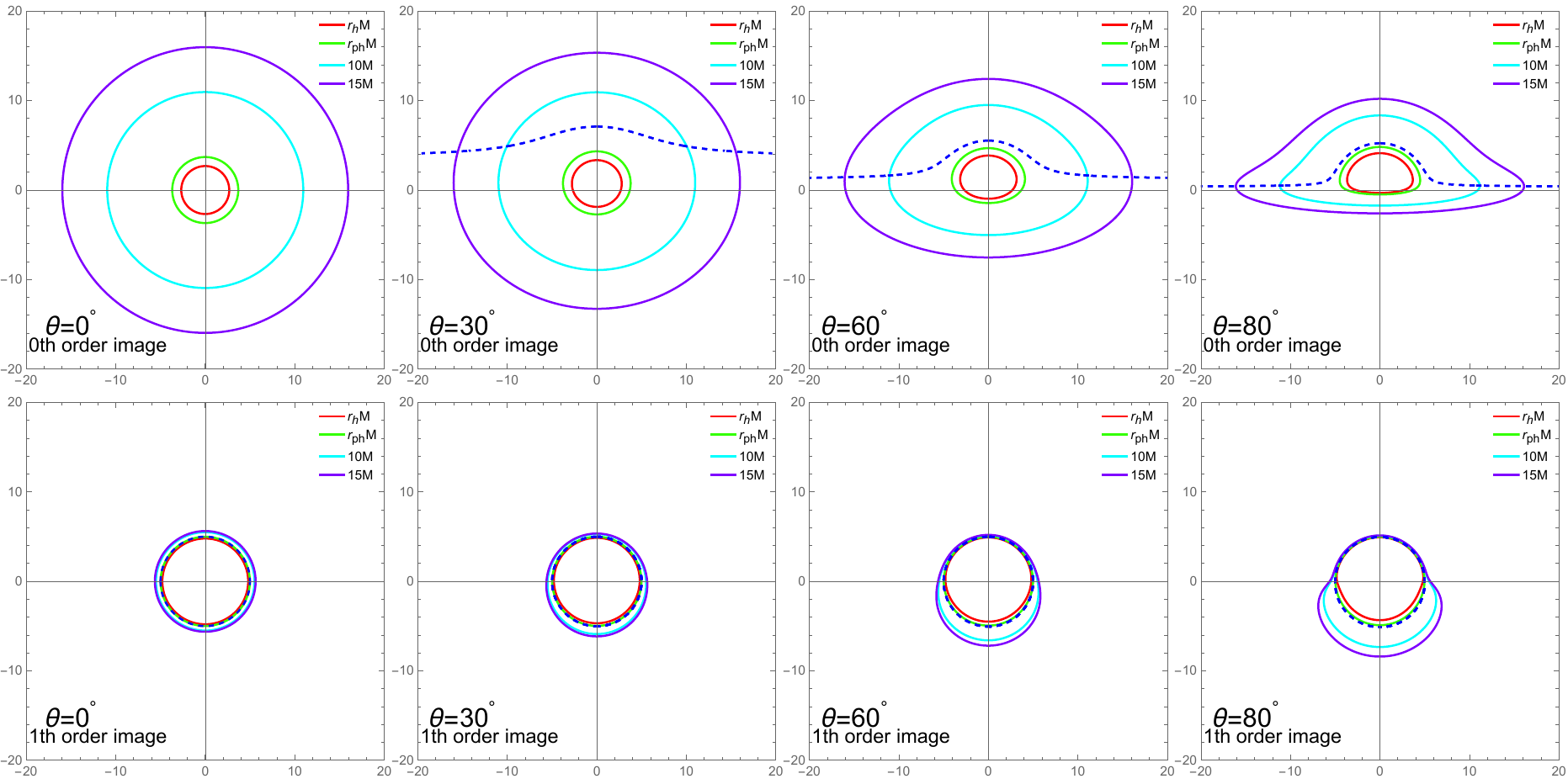}
		\caption{The $0^{\rm{th}}$-order image and the first image of accretion disk. The upper side is the $0^{\rm{th}}$-order image and the lower side is the first image. Blue dashed lines are perihelion curves, which intersect with images at $b_{\rm{max}}$. We set $q=0.5$, $a=1$ and $R=\rho=0.1$.}\label{fig_image}
		\includegraphics[width=1\textwidth]{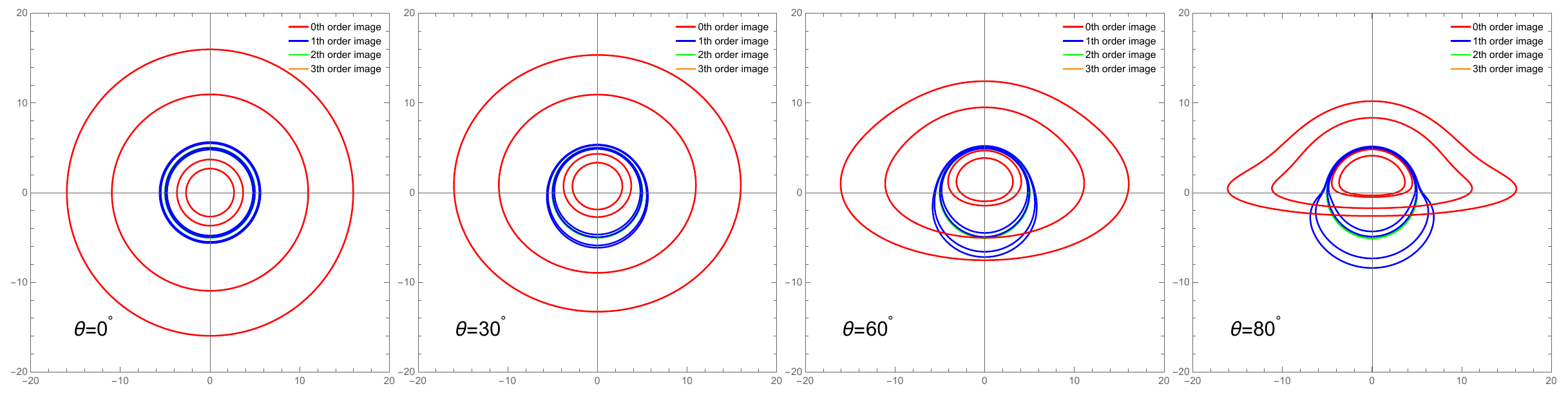}
		\caption{Image of thin accretion disk. The first four orders of images are plotted. The second images and the third images already can't be distinguished clearly in the figure. We set $q=0.5$, $a=1$ and $R=\rho=0.1$.}\label{fig_imagecombine}
	\end{figure}

		Now, the imaged of equal-$r$ orbit is plotted in Fig.~\ref{fig_image}. It is clear that figures agrees with our discussions above.
		
		At last, we merge the $n^{\rm{th}}$-order image ($n=0,1,2,3$) into Figure.~\ref{fig_imagecombine}.
	\newpage
	\begin{figure}[h]
		\centering
		\includegraphics[width=1\textwidth]{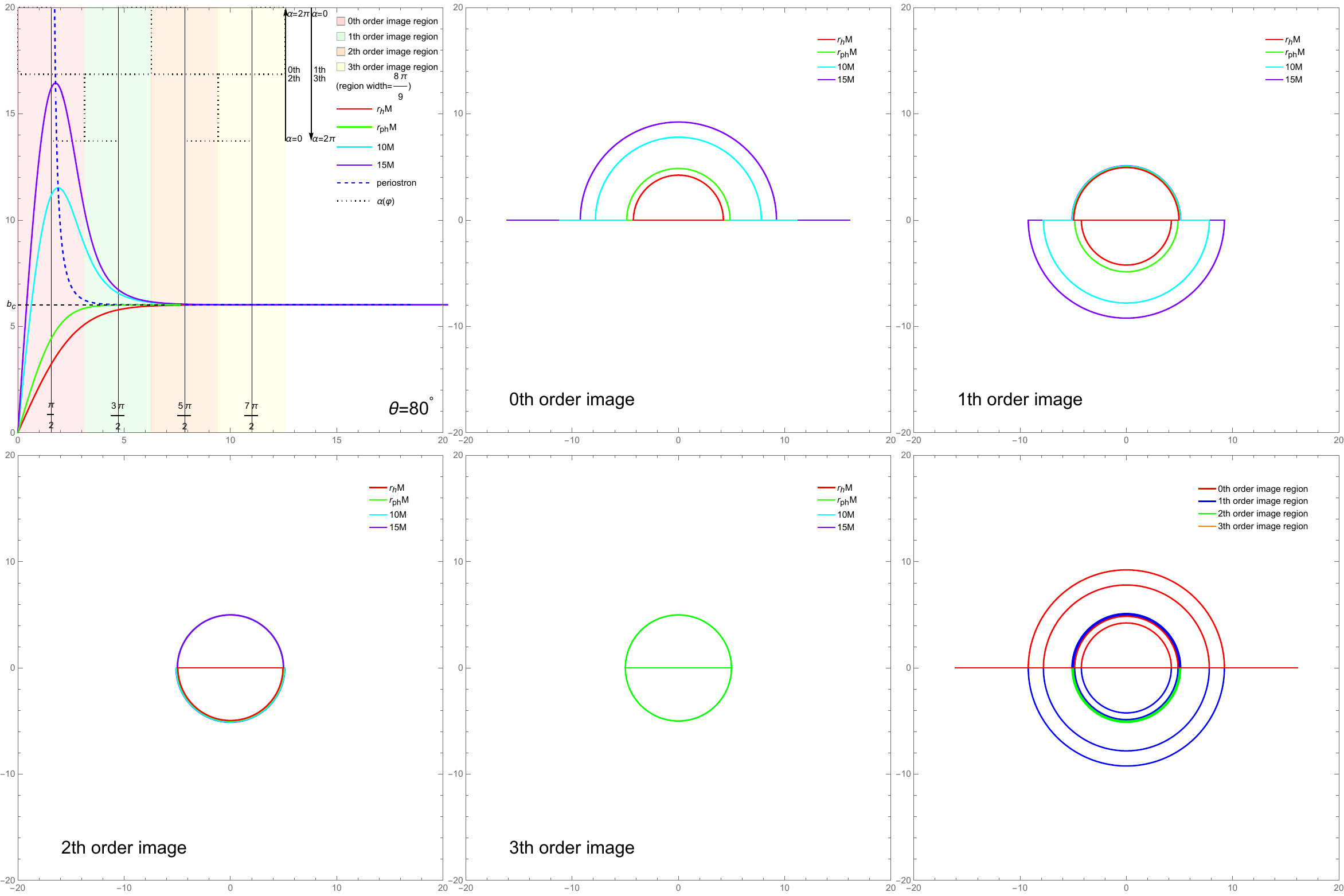}
		\caption{Image of thin accretion disk. The first four orders of images are plotted. We set $q=0.5$, $a=1$, $R=\rho=0.1$ and $\theta=\frac{\pi}{2}$.}\label{fig_90image}
	\end{figure}
	Here, we plot the image of equal-$r$ orbit for $\theta=\frac{\pi}{2}$ in Fig.~\ref{fig_90image}. When $\theta$ tends to $\frac{\pi}{2}$, S-shaped curves will become horizontal and vertical, which means that one could find arbitrary-order images of any source points on $\alpha=0$ and $\pi$ (X${'}$-axis). Of interest is that $\varphi\left(\alpha\right)$ will be a constant when $\alpha\in\left(0,\pi\right)$ and $\left(\pi,2\pi\right)$. It will cause a perfect circular images resulted by two source points $\phi=\frac{\pi}{2}$ and $\frac{3\pi}{2}$ in black hole's coordinate. Meanwhile, as shown in upper left corner of figure, the widths of $\varphi$ of images for any order will become $\pi$. This indicates that interval of $\varphi$ for neighbor-order images just link together. So, in the link region, for any $n$, the region $\alpha\in\left(0,\pi\right)$ of $n^{\rm{th}}$-order images will have a same shape as the region $\alpha\in\left(\pi,2\pi\right)$ of $\left(n+1\right)^{\rm{th}}$-order. It results in that the definition of image's order we have mentioned in Sec.~\ref{sec3_1} no longer works. Considering it is noticed that in Fig.~\ref{fig_imagecombine}, when $\theta=\frac{4\pi}{9}$, $0^{\rm{th}}$ image and the $1^{\rm{st}}$-order image for $r>r_{ph}$ both have a pit near X$'$-axis clearly. Reflected in Fig.~\ref{fig_1bphi}, it is due to that the colored lines ($b\left(\varphi\right)$ curves of equal-$r$ orbit) are steep in the region of $0^{\rm{th}}$-order images and $1^{\rm{st}}$-order images and S-shaped curves are horizontal. And as $\theta$ increases, these two pits will get closer to each other, and both will eventually arrive at $X'$-axis, leading to a series of circles appearing in Fig.~\ref{fig_90image}. We still draw the images of respective order in figure according this definition. 
	
\begin{figure}[htbp]
		\centering
		\includegraphics[width=1\textwidth]{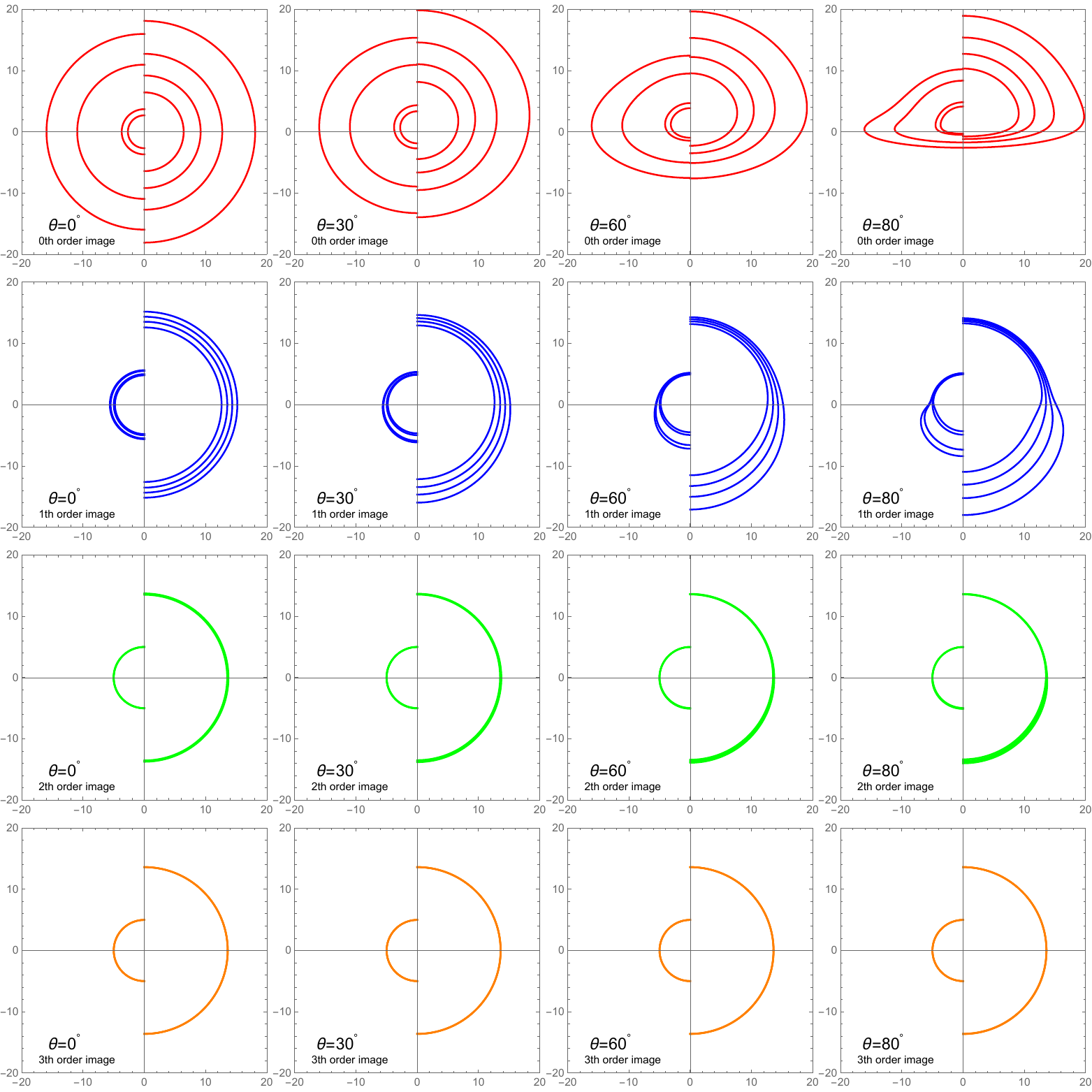}
		\caption{Comparation between two sets of images of thin accretion disk with different parameters. We set $q=0.5$, $a=1$, $R=\rho=0.1$ for the left side and $q=0.5$, $a=1$, $R=\rho=0.5$ for the right side. }\label{fig_imagecomparation}
	\end{figure}

	Here we give the comparation between two images with different parameters in Fig.~\ref{fig_imagecomparation}. As shown in the figure, black hole's size will  increase with the increase of CDM parameters $R$ and $\rho$.  When $\theta$ goes to $\frac{\pi}{2}$ and $r$ is sufficiently large, two sets of image will link with each other at $\alpha=\frac{3\pi}{2}$. It is because that one could find in Fig.~\ref{fig_1bphi}, colored lines near $\varphi=0$ will become extremely straight especially when $r$ is bigger. It causes that when $\theta$ goes to $\frac{\pi}{2}$, the left boundary of $\varphi$ of $0^{\rm{th}}$-order image will get close to the neighborhood of $\varphi=0$. Meanwhile, it is foreseeable that although $\theta$ is small, once $r$ is bigger enough, two sets of image could still link with each other at $\alpha=\frac{3\pi}{2}$.
	\newpage
	\subsection{Light intensity distribution on accretion disk}\label{sec3_3}
	\subsubsection{Novikov-Thorne model}\label{sec3_3_1}
		In this section, we use Novikov-Thorne model to study the image of thin accretion disk with brightness \cite{NT}:
	\begin{equation}\label{eq3_3}
		F\left(r\right)=-\frac{\mathcal{M}\Omega'}{4\pi\sqrt{-g}\left(E-\Omega L\right)^{2}}\int_{r_{isco}}^{r} \left(E-\Omega L\right)L'dr,
	\end{equation}
		where $F(r)$ is radiation flux emitted by time-like particle in circular orbit $r$, $\mathcal{M}$ is black hole's accretion rate, $g$ is metric
		determinant, and $r_{isco}$ is radius of ISCO we have introduced previously, $E$, $\Omega$, and $L$ are energy, angular velocity, and angular momentum of time-particle moving on the disk.
		\par
		Due to the different gravitational fields at disk and observer and their relative motion, the frequency shift will happen, the radiation flux observer received will be \cite{HY,F1,F2}
	\begin{equation}\label{eq3_4}
		F_{obs}=\frac{F\left(r\right)}{\left(1+z\right)^{4}},
	\end{equation}
		where $z$ is redshift factor defined as
	\begin{equation}\label{eq3_5}
		1+z=\frac{E_{r}}{E_{\infty}},		
	\end{equation}
		$E_{r}$ is the photon's energy received by an comoving observer with emitting particles passing through $Q\left(r,\frac{\pi}{2},\phi\right)$. When a photon propagates from the disk to infinity on the plane $\alpha/\alpha+\pi$, its energy received by the observer locating at $q\left(\infty,\theta,0\right)$ will be $E_{\infty}$. Photon's energy $E$ is the projection of photon's four-momentum $P_{\mu}$ onto the four-velocity $U^{\mu}$ of comoving observer:
	\begin{equation}\label{eq3_6}
		E=-P_{\mu}U^{\mu},
	\end{equation}
		Considering emitting particle moving in the circular orbit on the accretion disk, we get
	\begin{equation}\label{eq3_7}
		E_{r}=-\left(P_{t}U^{t}+P_{\phi}U^{\phi}\right).
	\end{equation}	
		Because observer at infinity only has a nonzero term $U^{t}=1$, we have
	\begin{equation}\label{eq3_8}
		E_{\infty}=-P_{t}. 
	\end{equation}
		So
	\begin{equation}\label{eq3_9}
		\frac{E_r}{E_\infty}=U^{t}\left(1+\Omega\frac{P_\phi}{P_t}\right),
	\end{equation}
		where $\Omega=\frac{U^{\phi}}{U^{t}}$ is angular velocity of emitting particle on accretion disk. And, as the projection of photon's four-momentum onto vector $\frac{\partial}{\partial \phi}$, $P_{\phi}$ is precisely the angular momentum $L$ of photon in black hole's coordinate system. So we have
	\begin{equation}\label{eq3_10}
		\frac{P_\phi}{P_t}=\frac{L}{-E}.
	\end{equation}
		To make our subsequent calculation, we transform $L$ into observer's coordinate system ($t',r',\theta',\phi'$), which takes photon's propagation plane $\alpha/\alpha+\pi$ as its equatorial plane and keep $t'=t$ and $r'=r$.
				
		Now, we give the projection of $\left(P_{t'},P_{r'},P_{\theta'},P_{\phi'}\right)$ onto vector $\frac{\partial}{\partial \phi}$. Due to $\theta'=\frac{\pi}{2}$, which means photon always move on its propagation plane $\alpha/\alpha+\pi$, $P_{\theta'}=0$. And it is obvious that basis vector $\frac{\partial}{\partial r'}$ is orthogonal with $\frac{\partial}{\partial \phi}$ at source point $Q\left(r,\frac{\pi}{2},\phi\right)$ that lead to zero contribution of $P_{r'}$. On the other hand, resulted by space-time itself is time-orthogonal ($\frac{\partial}{\partial t'}$ is orthogonal with $\frac{\partial}{\partial \phi}$), the contribution of $P_{t'}$ is zero. So, the upper-left corner of Fig.~\ref{fig_coordinate} will give the relationship between $\frac{\partial}{\partial \phi'}$ and $\frac{\partial}{\partial \phi}$ at source point $Q$, the angle $\eta$ between them is
	\begin{equation}\label{eq3_11}
		\cos\eta=\sin\theta\cos\alpha,
	\end{equation}
		which leads to relation between $P_{\phi}$ and $P_{\phi'}$
	\begin{equation}\label{eq3_12}
		P_{\phi}=P_{\phi'}\sin\theta\cos\alpha,
	\end{equation}
		where $P_{\phi'}$ is exactly photon's angular momentum $L'$ in observer's coordinate system. Further,
	\begin{equation}\label{eq3_13}
		\frac{P_\phi}{P_t}=\frac{L'\sin\theta\cos\alpha}{-E}.
	\end{equation}
		Please notice that in our coordinate setting, $L'<0$ for even-order images and $L'>0$ for odd-order images. And for odd-order images, Eq.~\ref{eq3_11} should be modified as $\cos\eta=\sin\theta\cos\left(\alpha+\pi\right)$. From above discussions, for any images of source point $Q$,
	\begin{equation}\label{eq3_14}
		\frac{P_\phi}{P_t}=\frac{\left|L'\right|\sin\theta\cos\alpha}{E}=b\sin\theta\cos\alpha,
	\end{equation} 
		where $b$ is impact parameter of photon shown in image point $q\left(b,\alpha\right)$.
		
		The four-velocity $U^{\mu}$ of observer comoving with a emitting particle moving in circular orbit on the accretion disk satify
	\begin{equation}\label{eq3_15}
		g_{tt}U^{t}U^{t}+g_{\phi\phi}U^{\phi}U^{\phi}=-1.
	\end{equation}
		Substitute $\Omega=\frac{U^{\phi}}{U^{t}}$ one could get
	\begin{equation}\label{eq3_16}
		U^{t}=\frac{1}{\sqrt{-g_{tt}-g_{\phi\phi}\Omega^{2}}}.
	\end{equation}
		According to Eq.~\ref{eq3_3} and Eq.~\ref{eq3_4},
	\begin{equation}\label{eq3_17}
		F_{obs}=\frac{-\frac{\mathcal{M}\Omega'}{4\pi\sqrt{-g}\left(E-\Omega L\right)^{2}}\int_{r_{isco}}^{r} \left(E-\Omega L\right)L'dr}{\left(\frac{1+\Omega b \sin\theta\cos\alpha}{\sqrt{-g_{tt}-g_{\phi\phi}\Omega^{2}}}\right)^4},
	\end{equation}	
		Take the derivative of $r$ for both sides of Eq.~\ref{eq3_15}, one could obtain
	\begin{equation}\label{eq3_18}
		g_{tt}'U^{t}U^{t}+g_{\phi\phi}'U^{\phi}U^{\phi}=0,
	\end{equation}
	which induce
	\begin{equation}\label{eq3_19}
		\Omega=\pm\sqrt{-\frac{g_{tt}'}{g_{\phi\phi}'}}.
	\end{equation}
		$\pm$ indicates two different directions of rotation for accretion disk and we will take positive in our next analysis. The energy $E$ and the angular momentum $L$ of emitting particle in Eq.~\ref{eq3_17} are 
	\begin{equation}\label{eq3_20}
		E=-g_{t\mu}U^{\mu}=\frac{-g_{tt}}{\sqrt{-g_{tt}-g_{\phi\phi}\Omega^{2}}},
	\end{equation}
	\begin{equation}\label{eq3_21}
		L=-g_{\phi\mu}U^{\mu}=\frac{g_{\phi\phi}\Omega}{\sqrt{-g_{tt}-g_{\phi\phi}\Omega^{2}}}.
	\end{equation}
	\subsubsection{Image of thin accretion disk with light intensity}\label{sec3_3_2}
	\begin{figure}[h]
		\centering
		\includegraphics[width=1\textwidth]{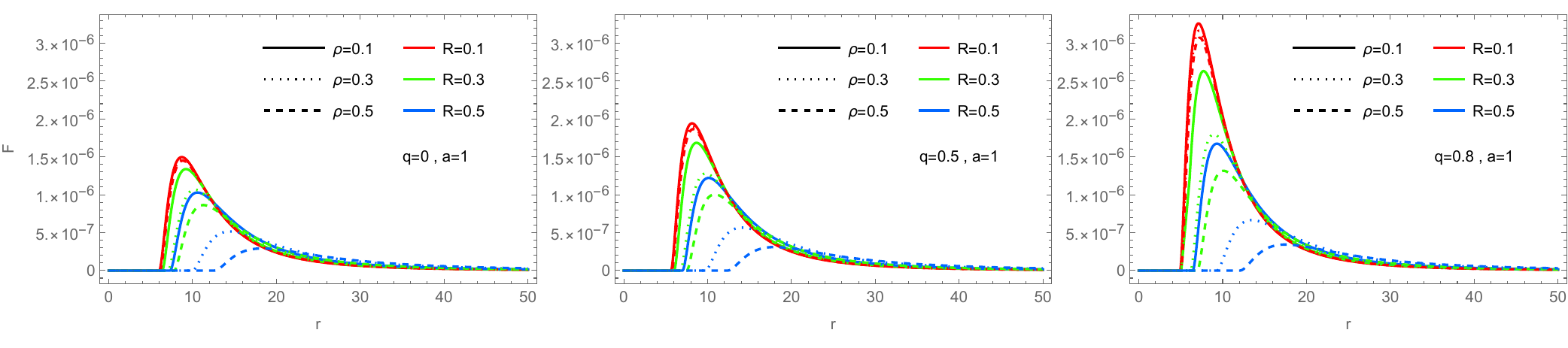}
		\caption{Radiation flux $F(r)$ as a function of $r$ for different parameters.}\label{fig_Fr}
	\end{figure}
	We give the figures of radiation flux $F\left(r\right)$ for different parameters as Fig.~\ref{fig_Fr}. As we can see in the figures, radiation flux will increase with the increase of $q$ and CDM parameters will constrain the radiation flux, especially $R$. Reflected in the figures, the peaks of blue dashed lines will become smaller as $R$ and $\rho$ increase.
	\begin{figure}[h]
		\centering
		\includegraphics[width=0.8\textwidth]{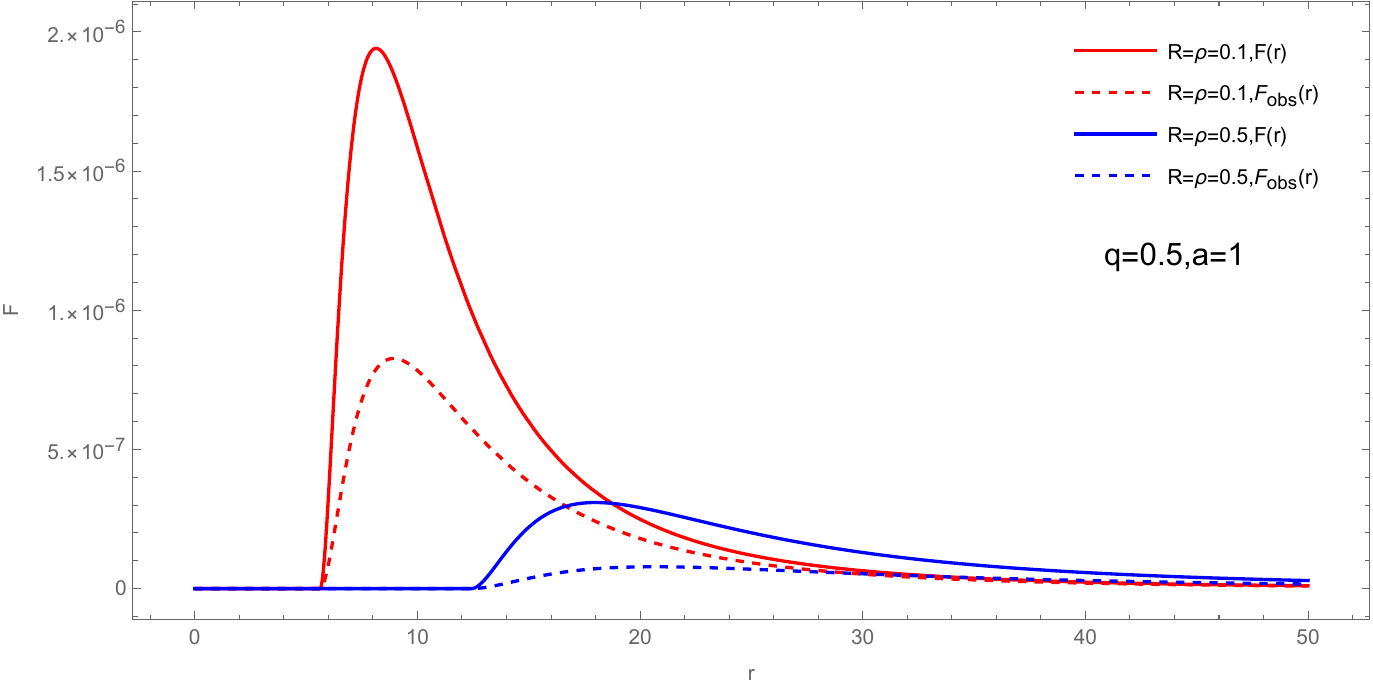}
		\caption{The comparison of radiation flux at light source  and observer when $\theta=0$.}\label{fig_Feo}
	\end{figure}
	
	Fig. \ref{fig_Feo} gave the comparison of radiation flux at light source  and observer when $\theta=0$. As one can see in the figure, gravitational red shift will significantly reduce the light intensity received by an observer in the distance. And with the increase of CDM parameter, this effect will become more distinct.
	
	According to the analysis above, we plotted the images of thin accretion disk with light distribution as Fig.~\ref{fig_4intensity}. It could be seen clearly:
	\begin{figure}[htbp]
		\centering
		\subfigure{
			\includegraphics[width=1\textwidth]{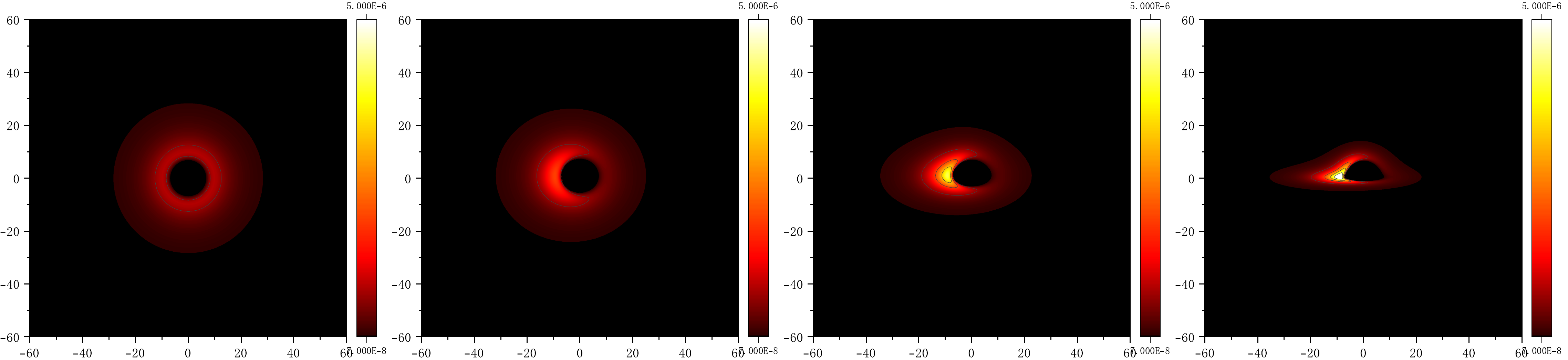}}
		\subfigure{
			\includegraphics[width=1\textwidth]{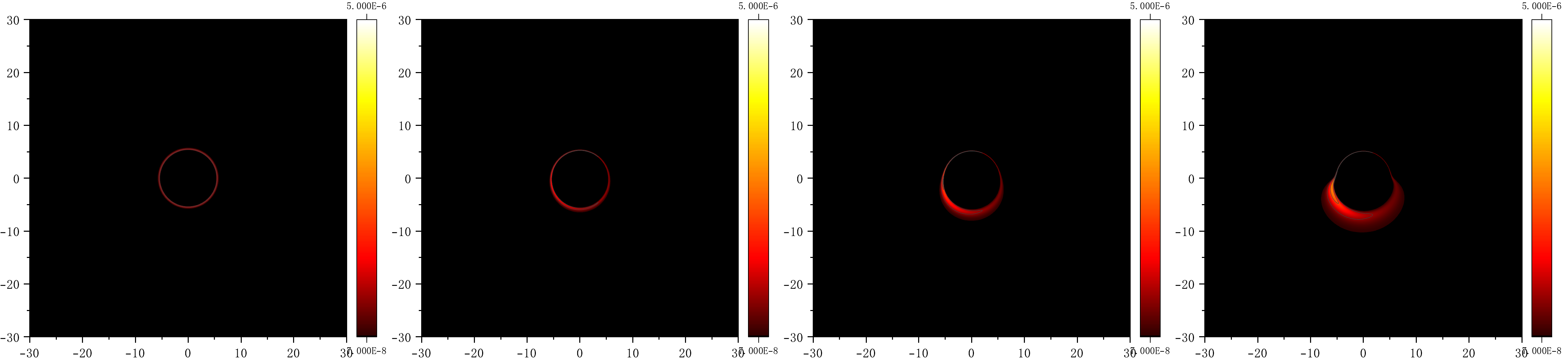}}
		\subfigure{
			\includegraphics[width=1\textwidth]{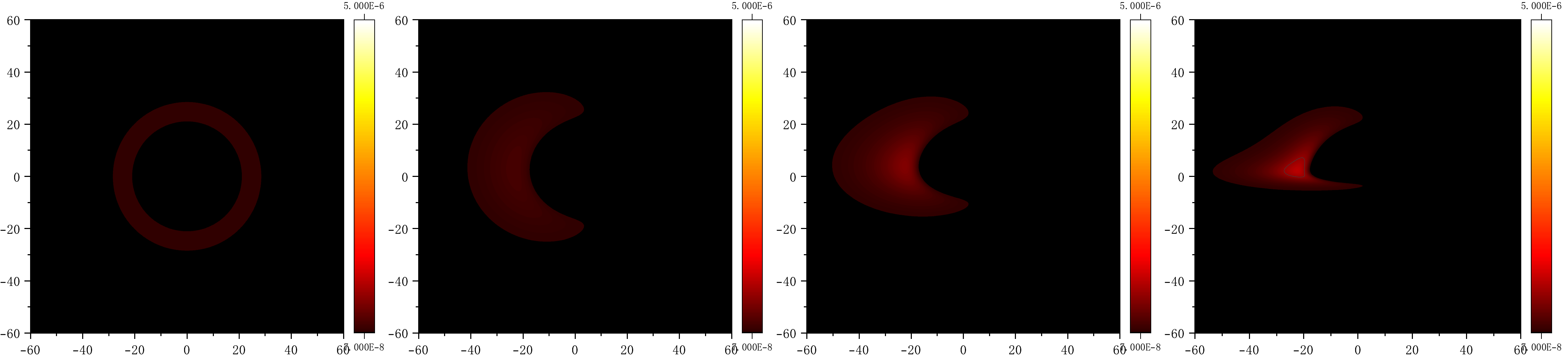}}
		\subfigure{
			\includegraphics[width=1\textwidth]{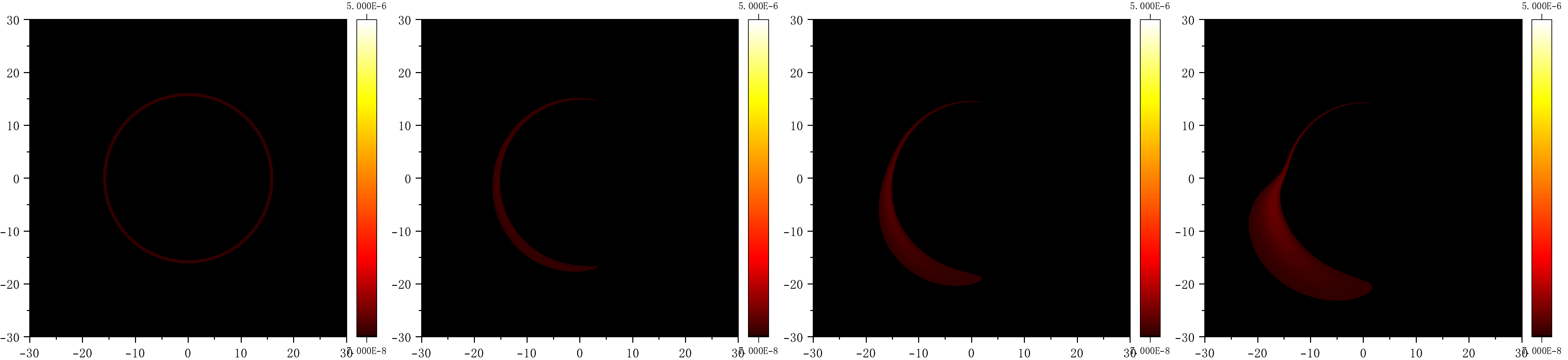}}
		\caption{The $0^{\rm{th}}$ image and the $1^{\rm{st}}$ images of accretion disk with light intensity. We set $q=0.5$, $a=1$, $R=\rho=0.1$ for the first two rows and $q=0.5$, $a=1$, $R=\rho=0.5$ for the rest. The first and the third rows are $0^{\rm{th}}$ images. The second and fourth rows are the $1^{\rm{st}}$ images. The inclination angles of observer are $0$, $\frac{\pi}{6}$, $\frac{\pi}{3}$ and $\frac{4\pi}{9}$ form left side to right side.}
		\label{fig_4intensity}
	\end{figure}
	\begin{itemize}
		\item [(a)] As the increase of inclination angle of observer, the left side of images become brighter and the right side dims. This is precisely attributed to the Doppler effect;
		\item [(b)] For the $1^{\rm{st}}$ images, the analysis in (a) still work. But it should be noticed that the intenser light in the left side of $1^{\rm{st}}$ image comes form the right side of accretion in fact;
		\item [(c)] CDM halo can increase the size of black hole and will vastly restrain the brightness of accretion disk.
	\end{itemize}
	
\begin{figure}[htbp]
		\centering
		\includegraphics[width=1\linewidth]{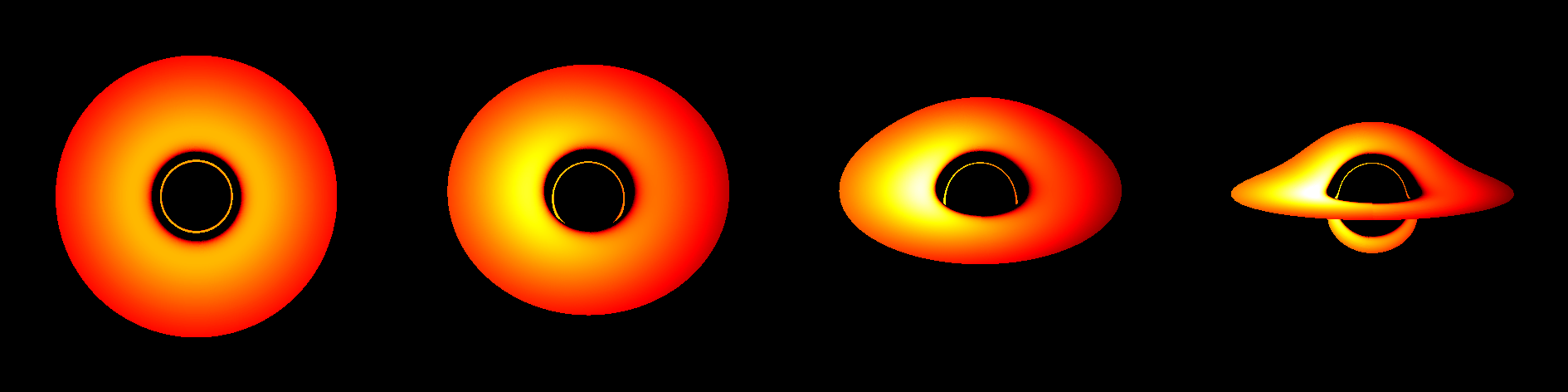}
		\caption{Actual images of thin accretion disk. We set $q=0.5$, $a=1$ and $R=\rho=0.1$.}
		\label{fig_intensity}
	\end{figure}

	At last, we choose a set of parameters and plot the actual images of thin accretion disk as Fig.~\ref{fig_intensity}. It is mainly formed by the $0^{\rm{th}}$ image and the $1^{\rm{st}}$ image factually.  
	\newpage
	\section{Conclusion and outlook}\label{sec4}
	We obtained a exact solution that describe EH black hole immersed in CDM halo and investigated its optical properties in detail. It is found in our research that CDM halo will significantly increase the size of black hole and reduce the radiation flux of thin accretion disk received by an observer at infinity. Besides, it is clear in article that the curve $b-\varphi$ will reveal the mechanism of thin accretion disk's imaging ($b$ is impact parameter of photon and $\varphi$ is deflection angle of photon when it arrives at circular orbit of time-like particles). Our research will provide a new view for subsequent study of image of black hole's accretion disk.
	
	In \cite{XZ}, a method was proposed to extend the static spherically symmetric black hole solution to rotating axisymmetric black hole solutions. Since real black holes often possess rotation, our next step could be to further investigate the rotating solutions of the EH black hole and explore their optical properties. Additionally, some ideas about spacetime symmetries emerged from the study in this article.
	\section*{Conflicts of interest}
	The authors declare that there are no conflicts of interest regarding the publication of this paper.
	\section*{Acknowledgments}
	We are grateful to Yu-Cheng Tang and Yu-Hang Feng for their useful suggestions. We also thank the National Natural Science Foundation of China (Grant No.11571342) for supporting us on this work.

	\bibliographystyle{unsrt}
	\bibliography{reff.bib}
\end{document}